\documentclass[12pt,aps,floatfix,prb]{revtex4}
\usepackage{amsmath}
\usepackage{latexsym}
\usepackage{float}
\usepackage{amssymb}
\usepackage{graphicx}
\usepackage{epsfig}

\begin{document}

\title{Coarse-grained models for fluids and their mixtures:
Comparison of Monte Carlo studies of their phase
behavior with perturbation theory and experiment}

\author{B. M. Mognetti$^{(a)}$, P. Virnau, L. Yelash, W. Paul, K. Binder \\
Institut f\"ur Physik, Johannes Gutenberg Universit\"at Mainz, \\
Staudinger Weg 7, 55099 Mainz, Germany \\
M. M\"uller\\
Institut f\"ur Theoretische Physik, Georg-August-Universit\"at
G\"ottingen\\
Friedrich-Hund-Platz 1, 37077 G\"ottingen, Germany\\
and \\
L. G. MacDowell \\
Dpto de Quimica Fisica, Facultad de Cc. Quimicas,\\
Universidad Complutense, 28040 Madrid, Spain}

\begin{abstract}
The prediction of the equation of state and the phase behavior of
simple fluids (noble gases, carbon dioxide, benzene, methane,
short alkane chains) and their mixtures by Monte Carlo computer
simulation and analytic approximations based on thermodynamic
perturbation theory is discussed. Molecules are described by
coarse grained (CG) models, where either the whole molecule (carbon
dioxide, benzene, methane) or a group of a few successive CH$_2$
groups (in the case of alkanes) are lumped into an effective point
particle. Interactions among these point particles are fitted by
Lennard-Jones (LJ) potentials such that the vapor-liquid critical point
of the fluid is reproduced in agreement with experiment; in the
case of quadrupolar molecules  a quadrupole-quadrupole interaction is
included. 
These models are shown to provide a satisfactory description of
the liquid-vapour phase diagram of these pure fluids. Investigations
of mixtures, using the Lorentz-Berthelot (LB) combining rule, also produce 
satisfactory results if compared with experiment, while in some previous
attempts (in which polar solvents were modelled without explicitly taking
into account quadrupolar interaction), strong violations of the LB 
rules were required. For this reason, the present investigation is a step
towards predictive modelling of polar mixtures at low computational
cost. 
In many
cases   Monte Carlo simulations of such models (employing the
grand-canonical ensemble together with reweighting techniques,
successive umbrella sampling, and finite size scaling) yield
accurate results in very good agreement with experimental
data.
Simulation results are quantitatively compared to an analytical
approximation for the equation of state of the same model, which is
computationally much more efficient, and some systematic discrepancies 
are discussed. These very simple coarse-grained models of small molecules
developed here should be useful e.g. for simulations of polymer solutions with such 
molecules as solvent.
\end{abstract}
\maketitle

 a) Electronic mail: mognetti@uni-mainz.de

\section{Introduction}
It has been a longstanding challenge to predict accurately the
equation of state and in particular the phase diagrams of fluids
and fluid mixtures from atomistic models via computer 
simulation.\cite{1,2,3,4,5} Such applications have required a widespread
development of computer simulation methodology: significant
advances were possible through the invention of Gibbs ensemble
\cite{6,7,8} and configurational bias \cite{9,10,11}
methodologies, grand canonical Monte Carlo simulations combined
with histogram reweighting methods \cite{12,13,14} and finite size
scaling \cite{15,16,17,18} including field mixing,
\cite{19,20,21,22} umbrella sampling \cite{23,24} and other
expanded ensemble methods.\cite{25,26,27} A lot of
effort has also been spent towards developing more and more accurate
effective potentials from quantum chemistry methods (e.g.
Refs.\ \onlinecite{28,29,30,31,32,33}). However, for simple 
and industrially relevant
fluids such as carbon dioxide \cite{34,35} it is still difficult
to predict the equation of state with high accuracy, such that
experimental data in the critical region and for temperatures $\pm
30 \%$ around it are reproduced to an accuracy of a few percent.
\cite{36,37} 
Extending such calculations to mixtures 
(in particular, solutions of polymers with supercritical carbon
dioxide as a solvent)
is even more
of a problem, due to the less complete knowledge of effective
potentials, and due to the extensive numerical effort required. A
three-dimensional parameter space involving the variables
temperature $T$, pressure $p$ and mole fraction $x$ needs to be
scanned for a binary system, and the phase diagrams are typically 
very complicated, because vapor-liquid and fluid-fluid phase
equilibria compete with each other.\cite{38,39,40,D_98}
If polymers are chosen as a solute, their molecular weight enters as
an additional fourth variable. Moreover, the coarse-grained representation
of the solvent (e.g.\ carbon dioxide) and the solute have to be 
compatible, i.e., one cannot combine an atomistic description of
the solvent with a much coarser representation of a macromolecular solute. 
There is clearly a need to
devise models that are simple enough to allow extensive simulation
studies with an affordable effort and nevertheless accurate enough
to be interesting for applications to experiment and in the
context of industrial processing. Such validated coarse-grained models that 
accurately reproduce thermodynamic bulk properties are also a starting
point for investigating the kinetics of phase separation or spatially
inhomogeneous systems (e.g.\ wetting and catalysis).

In the present work, we wish to make a step towards this goal,
extending our previous study of a selected sample of simple pure
fluids, in particular carbon dioxide \cite{36,37} to various
binary mixtures. 
We want to stress that our aim is not to reach the most accurate 
prediction of the phase diagram of a specific system. Indeed, 
motivated by the excellent results obtained for the pure carbon
dioxide and for simple quadrupolar molecules in general, \cite{36}
we want to investigate how this model performs for mixtures,
especially solutions of various alkanes. In particular
we will show that the new coarse grained (CG) model 
avoids the need for a big violation
of the Lorentz-Berthelot (LB) combining rules (that was required in previous work \cite{44}).
This violation
destroys the predictivity of the model because
extensive experimental data for the mixture would be required to determine
a parameter describing the violation of the LB combining rule.
Due to the generality of the approach and the level of accuracy 
for the pure components,
\footnote{
Indeed using results of Ref.\ \onlinecite{36} the CG parameters for  given
substances (with a reasonable quadrupolar moment) can be computed
in a straightforward way without additional simulations.  
}
the present investigation is relevant both for practical 
purposes and for a general understanding of coarse graining procedures.
\cite{Voth,Yip-2005,KT-2004}
We will also present results of an analytical Equation of State (EOS)
which (apart from some region of the phase diagram near critical points)  is
able to yield  
rather satisfactory predictions in agreement with Monte Carlo results.
It is very important to note that this EOS uses the same model parameters 
as the Monte Carlo simulation. This implies that in principle we
are in a position to attempt to predict the phase diagram of a binary
mixture (which is very complex \cite{38,39,40,D_98}) with comparatively small
computational effort. In this view the reader should also interpret 
our choice to use LB combining rules: of course there are no reasons
 to believe
that such approximations should be exact,
and certainly there will be cases where more complicated combining
rules are preferable. However, the simple LB combining rules used here
suffice for a wide class of systems with quite acceptable errors.

Due to the generality of the scheme presented in this work we expect
discrepancies, and some regions of the phase diagram might not be 
predicted properly. This is related to several limitations of the present
procedure like {\em i}) the large $T$ expansion involved in the building
of the CG model for quadrupolar solvents \cite{36,37} and {\em ii}) limitations
related to our simple modeling approach like the simple potentials involved (Lennard-Jones),
the neglect of atomistic details, and the use of the LB combining
rules for which discrepancies are 
\footnote{
However sometimes is far from being clear if deviations from LB combining
rules are infact compensation of some bias of the model used. For instance in
the case of polar substances the present work  shows very clearly how 
strong violations of the rules in Ref.\ \onlinecite{44} were more properly 
due to a bad modelling for the solvents. 
}
 known to arise (see e.g.\ \onlinecite{Hicks77} for some systems 
also investigated in
this work). In order to disentangle point {\em i}) from point {\em ii}) we also
present investigations of similar apolar mixtures for which the new  
CG model \cite{36,63} does not result in any improvement. The results 
show similar discrepancies from experiment as the polar phase diagrams, 
confirming the quality of the choice in Refs.\ \onlinecite{36,63}. 
We want to stress that in order to test the goodness of
our CG model, the only reliable method is a Monte Carlo
investigation. 
Indeed, without MC simulation it is impossible to distinguish 
the bias related to the approximations involved in the EOS from 
the bias involved in the CG model [point {\em ii}) above]. For instance, 
we will present results for the mixture of methane and carbon dioxide 
for which EOS results will be in better agreement with experiments 
than MC results: this is clearly a fortuitous cancellation!

It is important to report that other interesting and significant attempts 
to build a systematic description of mixture phase diagrams are present in
the literature. For instance in \onlinecite{Stoll2003,Vrab2005} 
mixtures are treated
with models previously investigated in \onlinecite{Vrab2001}.
For some of the molecules studied, these models allow for an additional
parameter that can be adjusted and consequently a more accurate fit
of experimental data is possible.
On the other hand, there is a loss in predictivity because
the full phase diagrams of the pure substances are required
in order to determine 
the simulation parameters (computed in a $\chi$ square fit which minimizes
discrepancies with experiment\cite{Vrab2001}) plus mixture data
\cite{Stoll2003,Vrab2005} to determine the mixing parameters. 
\footnote{
It is interesting to observe how atomistic models that are not improved 
to describe all the pure substances phase diagram but take as
experimental input only the critical temperature and density (like in our case)
are less accurate than our simpler (see for instance the discussion of
EPM2 model \cite{HY-95} in \onlinecite{36}).
}
So the strategy of the present work is to deal with relatively simple models,
where (in the framework of Monte Carlo simulations) the statistical mechanics
can be dealt with at a very good level of accuracy (e.g.\ long runs employing
advanced Monte Carlo techniques are possible to minimise statistical 
errors and systematic errors
due to finite size effects which are avoided by finite size scaling
analysis). These 
models are suitable for analytic EOS models as well, and can serve as a 
starting point for the coarse grained modeling of polymer solutions. Of
course, we do not imply that a complementary simulation strategy (making
models as detailed as possible, to account for the packing of molecules in
the liquid as accurately as possible, including polarizability, etc.) is
not worth pursuing in its own right, but it is outside of the scope of the
present work.

\section{Computational details and outline
}

It is well established
\cite{5,19,20,21,22,41,42,43,44,44a,44b} that the most reliable approach to
study the phase behavior of fluids is based on grand canonical
Monte Carlo simulations together with histogram reweighting and finite size
scaling techniques, especially if one wishes to include the critical region. In this
study, we follow this approach, and amend it by successive
umbrella sampling \cite{24} to obtain coexistence curves far from
criticality. This method has the additional advantage that the
interfacial free energy between the coexisting phases can be
extracted as well.\cite{47,48,49,50} As we are interested in a very
fast simulation code, we omit any potentials including effective
charges, and restrict our attention to short range effective
potentials. Three-body (nonbonded)
forces are avoided as well. Electrostatic quadrupole-quadrupole interactions
are treated as a perturbation (which is practically justifiable
\cite{37}), such that an effective angular-independent (but
temperature-dependent \cite{36,37,51,52}) interaction decaying
proportionally to the power $r^{-10}$ of the interparticle distance
$r$ results. The dispersion forces are modeled by Lennard-Jones
(LJ) potentials. For the sake of computational efficiency, all
potentials are cut at the distance $r=r_c=2(2^{1/6})\sigma$ and
shifted to zero at r$_c$ ($\sigma $ is the range parameter of the LJ
potential). When we deal with alkane chains, we disregard any
torsional forces and bond-angle potentials and integrate a few
successive chemical monomers into one effective monomeric unit
(cf.\ fig.~\ref{fig1}). This is done in the way that one such unit contains
three carbon-carbon bonds between successive carbon atoms, and we
do not distinguish between interior CH$_2$ monomers and the CH$_3$
groups at the chain ends. Thus, for example, hexadecane
(C$_{16}$H$_{34}$) is represented by a chain molecule containing
five effective monomers (see fig.~\ref{fig1}).\cite{44,51} 
The procedure of coarsening 
three carbon atoms in a bead has been proven to be optimal
in several theoretical investigations  \cite{54} 
(see Sec.\ 4.3.2). We stress that the particular choice of coarsening
three carbon units into one bead has nothing to do with the physical lengths
of the chain (like for instance the Kuhn length), but is a choice 
that depends more on the potentials used. Indeed as
neighboring beads along a chain interact with a bonding potential 
(see Sec.\ IIIC for definitions) in addition to the Lennard Jones potential, the
coarse grained model of the chain exhibits a degree of local stiffness, although neither
bond angle or torsional potentials are included explicitely.
This implies that the Kuhn length is longer
than the diameter of our beads.

Of course, the suitable choice of parameters is crucial for such coarse-grained models:
we choose the strength of the quadrupole moment $Q$
(if there is one) such that it is compatible with experimental data, and adjust
the range $\sigma$ and strength $\epsilon$ of the LJ potential
such that the experimental critical density $\rho_c$ and critical
temperature $T_c$ are reproduced precisely in the simulation. In
Sec.\ III, we will briefly discuss the accuracy of this procedure
for a variety of pure systems (noble gases, CO$_2$, CH$_4$,
C$_6$H$_6$, short alkanes) while Sec.\ IV contains the central
part of our work, in which we present a variety of results for binary
mixtures. The additional interactions needed for the mixtures are
chosen by the simple Lorentz-Berthelot combining rules.\cite{52}

Technical aspects of our simulations are similar to previous 
studies.\cite{36,37,44} Far from the critical point coexistence
densities are computed using the successive sampling algorithm of 
Virnau and M\"uller \cite{24} in which high free energy barriers 
are overcome constraining the algorithm --at a certain time of the
simulation-- to sample configurations of a system where the
number of particles is $n$ or $n+1$.
Varying  $n$  from $n=0$ to $n=N_\mathrm{MAX}$ one is able to reconstruct (after
proper reweighting) the free energy profile $F(n)$ at coexistence in the range 
of densities of interest. 
At phase coexistence, we expect a distribution $F(n)$ with two
peaks (corresponding to the two coexisting phases that differ in 
particle number) which have equal weight.
In few very fast runs (using a small cubic box
$L\approx 7 \sigma_M$, where $\sigma_M$ is the biggest LJ length parameter of
the model), invoking the equal weight rule for $F(n)$, we
are able to tune the chemical potential(s) to their coexistence values, with
a reasonable  error ($\approx$1-5\%) which in some cases should be  enough.
Then, we start a second long simulation for a larger elongated box (to
enhance the formation of the liquid-gas interface) $V=2\cdot L^3$ with $L=9
\sigma_M$ in which every window is sampled with 2-10$\cdot 10^4$ MC 
steps. Every MC step includes: 100 grand canonical moves in which we try to 
 insert/delete solvent (and chain) particles, 1 local move in which a number
 of monomers equal to the total number of monomers are rearranged, and
 10$\cdot$N$_\mathrm{chain}$ reptation moves, where   N$_\mathrm{chain}$ is the
 number of the chains in the box.
Such a run requires on average 10 h of cpu time on 32 nodes of an IBM Power4
cluster.  The precision of the measured coexistence densities (for
instance) is roughly 1\%. Using a spherical averaged potential allows us to speed
up 
computations by a factor $\approx$5 \cite{37} in comparison with the full quadrupolar model. A number of chains 
 $N_\mathrm{MAX}\approx 1100$ usually allows a complete sampling of 
the liquid peak, while the number of  solvent particles is 
typically of the same order of magnitude. We emphasize that --unlike simulations
in the Gibbs-ensemble-- in addition to the densities and compositions of
the coexisting phases and their compressibilities, the simulation technique
also provides information about the interface tension. 
 At the critical point,
we use the same kind of simulation described above, but unconstrained. (At
every time the number of particles is free to fluctuate in all the region
[0,$N_\mathrm{MAX}$]). For more detail on the finite size analysis used
we refer the reader to Sec.\ IVC.

Even with all these simplifying approximations, establishing the
phase behavior and thermodynamic properties of binary mixtures
comprehensively still requires a lot of work with Monte Carlo
simulations. Far away from critical points, such an effort is not
needed, and one can try to use an analytical equation of
state. We use a previously developed theory based on 
Wertheim thermodynamic perturbation theory 
\cite{52´} (TPT). We strictly follow Ref.\ \onlinecite{53,McD_2002}. 
In particular the free energy of the system $A$ is decomposed in
a contribution due to a mixture of unbonded monomers (the reference system) 
plus a contribution due to chain associativity $A_\mathrm{chain}$,
\cite{McD_2002} $A=A_\mathrm{ref}+A_\mathrm{chain}$. Wertheim's 
theory allows us to compute $A_\mathrm{chain}$ perturbatively 
using quantities of the reference system (like pair correlation functions)
and the known bonding potential. We use a first order perturbation
theory (TPT1) which (at this point) reduces the problem to the 
computation of pair correlation functions and the free energy
($A_\mathrm{ref}$) 
of a  binary mixture of  non-bonded monomers interacting 
with LJ potentials (chain-chain monomers and solvent-chain monomers) and the LJ
+ quadrupolar interaction (see Sec.\ IIIB) for solvent-solvent monomers.  
$A_\mathrm{ref}$ is computed using standard perturbation theory: the
Ornstein-Zernike equation is solved using a Mean Spherical (MSA)
closure.\cite{HM-86}
 In particular, one chooses as reference system a mixture of
hard spheres with diameters computed using the repulsive part of  the
monomer-monomer  potential in a Barker Henderson approximation, \cite{BH-67}
while the attractive part of the potential is treated as a perturbation.
A MSA solution is then obtained using the  analytical implementation of 
Tang and Lu, \cite{53a,53b} in which the repulsive part of the LJ
potential is fitted by a couple of Yukawa tails which allow to
obtain an analytical result.\cite{McD_2002} In our present
modeling approach we need to consider LJ potentials plus quadrupolar
interactions. This problem has been solved in \onlinecite{36} (see App.\ A
of \onlinecite{36}) by applying a second pair of Yukawa tails to fit the 
quadrupolar interaction. 
In our MSA scheme we also use a ``one fluid approximation''.\cite{HM-86}
Our results will show how this simple theory is able to reproduce
 results in rather good agreement with MC data away from the critical point.
On the other hand, big discrepancies occur near the critical points, due
to the Mean Field nature of the MSA while experimental results exhibit
critical behavior characteristic of the Ising universality class. 
We are aware of significant efforts 
 to design proper EOS which include Ising fluctuation near the 
critical point.\cite{Parola_2008,Parola_1984} However, such
investigations are beyond the scope of the present paper.
Another popular method based on TPT1 is known  as 
``statistical associating fluid theory'' (SAFT). \cite{saft}

We want to stress that Monte Carlo simulations
remain an indispensable tool in investigations of the phase behavior 
of polymer solutions and mixtures. 
Indeed,
in the present study the model parameters 
($\epsilon$, $\sigma$ and $q_c$) have been determined\cite{36} 
using the simulation critical points which were obtained by Monte 
Carlo simulation in \onlinecite{36}. (Any mean field approximation
has difficulties in reproducing the critical line with sufficient accuracy). 
Note that supercritical fluids are interesting
and useful for practical applications, mainly
due to 
their high compressibility and the concomitant large variations
of density upon small changes of pressure,
which are the origin of the breakdown of a mean field
approximation like TPT. 
 This means that in some very interesting regions of the phase
diagram  Monte Carlo simulations are indeed a very valuable tool.

\section{Phase behavior of selected pure systems}

When we discuss the extent to which the Lorentz-Berthelot combining
rule can account for the phase behavior of mixtures, we need to
distinguish between inaccuracies arising from an imperfect
description of the pure components and those arising from the
Lorentz-Berthelot rule.
Therefore it is
necessary to give an overview of our modeling of the pure
components at the outset. Note that a possible additional source of errors
are entropic packing effects of non-spherical molecules that may
show up differently in a mixture of two molecules having different
shapes rather than for a pure system, where all molecules have the
same shape. Such effects are lost in our coarse-grained
models. However, this latter criticism cannot be applied when we
consider mixtures of noble gases, since in the framework of
classical statistical mechanics the description of noble gas atoms
as point particles, where two such atoms interact with a potential
depending on the absolute value of their distance only,
is certainly appropriate. (Disregarding the case of He, quantum effects
are negligible indeed at temperatures of interest \cite{55}).
For that reason, noble gases are also included in our discussion,
because they will bring out the possible limitations of our
modeling in terms of pair-wise effective potentials between
point-like particles most clearly. Thereafter, we shall deal with
CO$_2$, C$_6$H$_6$, CH$_4$, and selected short alkanes.

\subsection{Noble Gases}
The interaction between neutral point-like particles in our work
is always described by the Lennard-Jones (LJ) potential,

\begin{equation}\label{eq1}
U_{ij}^{LJ} = 4 \epsilon [(\frac {\sigma}{r_{ij}})^{12} -
(\frac{\sigma}{r_{ij}})^6]\quad .
\end{equation}

Rather than working with the full LJ potential as written in
Eq.~(\ref{eq1}), we find it computationally more convenient and
efficient to cut off this potential at $r=r_c=2^{7/6} \sigma$ and
shift it to zero there, such that

\begin{equation}\label{eq2}
U_{ij}(r) = U_{ij}^{LJ}(r)+ 4 \epsilon S\quad , \quad U_{ij}(r\geq r_c)=0\quad ,
\end{equation}

where S = 127/16384 for our choice of $r_c$, so that the potential
is continuous everywhere. When we require that Eqs.~(\ref{eq1},
\ref{eq2}) yield a vapor-liquid phase diagram such that the
critical temperature $T_c$ coincides with the experimental
critical temperature $T^{\textrm{exp}}_c$ of a particular system,
the strength ($\epsilon$) of the LJ potential is fixed once
$T_c^* = k_BT_c/\epsilon$ has been determined for the model.
Likewise, requiring that the critical density $\rho_c$ of the
model coincides with the experimental critical density
$\rho_c^{\textrm{exp}}$ of that system the range $(\sigma)$ of the
LJ potential is fixed once $\rho_c^*=\rho_c \sigma ^3$ is known for the
model. Here, $T^*=k_BT/\epsilon $ and $\rho^*=\rho\sigma
^3$ are dimensionless temperature and density, respectively.
Actually, the phase diagrams of both the full (untruncated) LJ
potential and of its truncated version Eqs.~(\ref{eq1}, \ref{eq2})
have been estimated with high precision.\cite{44,49}
Fig. 11 of Ref.\ \onlinecite{36} compares these phase diagrams 
with each other and
with experimental data for the noble gases Ne, Ar, Kr and Xe.
\cite{56} One can see that in this scaled representation the
differences between the phase diagrams based on full and truncated
LJ models are quite minor. Although noble gases are thought to be
the best possible experimental realization of LJ fluids, the
agreement is not perfect either: while Ne and Ar are very close to
the LJ prediction, the data for the fluid branch of Kr and Xe are
somewhat off. This implies that even noble gases do not strictly
satisfy the ``law of corresponding states'', and hence a
description in terms of classical point particles interacting with
purely pairwise potentials of the same functional form,
$U_{ij}(r)=\epsilon \; f(r/\sigma)$, with one parameter for the
strength $(\epsilon$) and another for the range $(\sigma)$ of the
potential cannot be strictly true, irrespective of the form of the
function $f(r/\sigma)$: either a more complicated form of the
pairwise interaction, involving a third system-specific parameter
is needed, or (what is usually assumed) some effect of three-body
interactions \cite{57,58,59} are present.

An even more pronounced deviation from the simple LJ model shows
up, however, when additional quantities are analyzed, such as the
vapor pressure $p_{\textrm{coex}}(T)$ at liquid-vapor coexistence
and the interfacial tension $\gamma(T)$ between the coexisting
vapor and liquid phases of the fluid (see figs.~\ref{fig3},
\ref{fig4}). It is clear that adjusting $\sigma$ from
$\rho_c^{\textrm{exp}}$ implies that the whole curve for the
coexistence pressure $p_{\textrm{coex}}(T)$ in the $(p,T)$ plane
is underestimated for both Kr and Xe. This is a serious drawback
for the description of binary mixtures, of course, where one
wishes to work in the $(T,p,x)$ ensemble, $x$ being the molar
fraction of the solute. Therefore, we have tried an alternative,
namely adjusting $\sigma$ such that the experimental critical
pressure $p_c^{\exp} =p_{\textrm{coex}}(T_c)$ is correctly
reproduced. For Kr the critical temperature $T_c^{\exp} = 209.46~K$ 
\cite{56} implies $\epsilon = 2.8971 \cdot 10^{-21}~J$. If one
uses $\rho_c=11.0~\mathrm{mol}/\ell$  \cite{56} to fit $\sigma $ one
obtains $\sigma = 3.6524~$\AA , while using $p_c^{\exp}= 55.20$ bar
\cite{56} instead would yield $\sigma = 3.58782~$\AA.
(For a discussion of the accuracy of our estimation of $\epsilon$ and
$\sigma$, we refer to table I. In order to guarantee the 
reproducibility of our results we always present $\epsilon$ and 
$\sigma$ with all the digits that have been used in our programs.)
Fig.~\ref{fig4} shows that a somewhat better description of
the vapor pressure $p_{\textrm{coex}}(T)$ is obtained over the
full temperature regime from $140  K < T <T_c^{\exp}$. 
The deviation from the data for the surface tension $\sigma$ has also
become smaller (fig.~\ref{fig3}b), but now there is a strong
deviation between the data for the liquid branch of the
coexistence curve and the model (fig.~\ref{fig3}a). 
Similar problems are observed for
Xe, where $T_c^{\exp} = 289.74K$ yields $\epsilon =
4.00747 \cdot 10^{-21}J$, while the use of $\rho_c^{\exp}=8.371~\mathrm{mol}/\ell$
yields $\sigma = 4.00053 \mathrm{\AA}$ and use of $p_c^{\exp}= 58.41$ bar
yields $\sigma = 3.92326 \mathrm{\AA}$. Figs.~\ref{fig3}, \ref{fig4} show
that for these noble gases the description of the coexistence
curve, vapor pressure at coexistence and surface tension
is clearly not as good as for the model of CO$_2$ and C$_6$H$_6$ proposed in
Ref.\ \onlinecite{36}. These problems carry over to our modelling of
binary rare gas mixtures (see Sec.~III A), as the comparison with
experimental data shows.\cite{59'}
At this point we recall that as outlined in the introduction,
the investigation of such a system has been undertaken in order
to get an order of magnitude estimate of the errors inherent to
our very simple models: the goal is not the derivation of a very 
elaborate description of noble gas mixtures.
Figure \ref{fig4} clearly shows that this model allows for a fairly good
description of the mixture phase diagram (if compared to other
mixtures presented in this work). In the present 
context it was not necessary to include more complex potentials 
available in the literature since long
times (e.g.\ \onlinecite{ARD-79,KH-70,HK-72}).

\subsection{Small Molecules: Methane, Carbon Dioxide, Benzene}
Methane (CH$_4$) is also described as a point particle, and again we
take Eqs.~(\ref{eq1}), (\ref{eq2}) as a coarse-grained
description of the interaction between methane molecules. Using
$T_c^{\exp}= 130.6~K$ \cite{56} and $\rho_c^{\exp} = 10.1~\mathrm{mol}/\ell$ 
\cite{56} as experimental input to determine
$\epsilon$ and $\sigma$, we obtain $\epsilon =
2.63624\cdot 10^{-21}~J$ and $\sigma = 3.75792~\mathrm{\AA}$. Fig.~\ref{fig5}
compares the resulting model prediction for the coexistence curve
in the temperature-density plane, the vapor pressure at
coexistence and the surface tensions with the corresponding
experimental data.\cite{56} It is remarkable that in this case
the simple potential model \{Eqs.~(\ref{eq1}), (\ref{eq2})\} works
better than in the case of the noble gas.

For molecules such as carbon dioxide (CO$_2$) and benzene
(C$_6$H$_6$) the situation is more complicated: while CH$_4$ is a
molecule of approximately spherical shape and does not have a
quadrupole moment, both CO$_2$ and C$_6$H$_6$ have quadrupole
moments. Note that (at least to a very good approximation
\cite{60,61}) CO$_2$ is a linear molecule while C$_6$H$_6$ is
disk-like. In \onlinecite{36} we have shown that a very good description
for both molecules is obtained when Eqs.~(\ref{eq1}) are augmented,
(\ref{eq2}) by a quadrupole-quadrupole interaction term. As the
latter is only a relatively small perturbation of the
Lennard-Jones-type interaction, it suffices to treat the
(angular-dependent) quadrupolar interactions via thermodynamic
perturbation theory. To leading order this yields the
following effective potential \cite{36,62,63}

\begin{equation}\label{eq3}
U_{ij}^{IQQ} = - \frac 7 5 \frac {1}{k_BT} Q^4/r_{ij}^{10} .
\end{equation}

Here, Q is the strength of the quadrupole moment of the considered
molecule. Note that the interaction is isotropic and inversely
proportional to temperature. We also cut off this part of the
interaction at the same radius $r_c$ as the LJ interaction, and
shift it to zero at r$_c$ as well, which yields the following total
pairwise interaction for these molecules

\begin{eqnarray}\label{eq4}
U(r_{ij})= \left\{ \begin{array}{r@{\; , \;}l} 4 \epsilon
[(\sigma/r_{ij})^{12}-(\sigma/r_{ij})^6- \frac{7}{20}
q(\sigma/r_{ij})^{10} + S] & r \leq r_c \\ 0 & r \geq r_c
\end{array} \right.
\end{eqnarray}

where

\begin{equation}\label{eq5}
S = \frac {127}{16384} + \frac 7 5 \frac {q}{256}
\end{equation}

and $q$ is the reduced quadrupolar interaction parameter,

\begin{equation}\label{eq6}
q = Q^4/[\epsilon \sigma ^{10} k_BT] = q_cT_c/T\quad , \quad
q_c\equiv q(T_c) \quad .
\end{equation}

Note that Eq.~(\ref{eq6}) is given in CGS units; in SI units, there
would be an additional factor $(4\pi \epsilon_0)^{-2}$.

Using Eqs.~(\ref{eq4})-(\ref{eq6}), one can fix $\epsilon$
and $\sigma$ such that critical temperature $T_c^{\exp}$ and
density $\rho _c^{\exp}$ are reproduced. (For $Q$,
the experimental value is  taken as a first guess). As discussed in
Ref.~\onlinecite{36}, this leads to a self-consistency problem, since
Eq.~(\ref{eq6}) must hold together with

\begin{equation}\label{eq7}
\epsilon(q_c) = k_BT_c^{\exp} /T_c^* (q_c), \quad \sigma ^3 (q_c)
= [ \frac {\rho_c^* (q_c)M_{\textrm{Mol}}}{\rho_c^{\exp}N_A}] \;,
\end{equation}

where $M_{\textrm{Mol}}$ is the molar mass of the molecule and
$N_A$ is Avogadro's number. This problem was solved in \onlinecite{36}
by determining the functions $T_c^*(q_c)/T_c^* (0)$, and $\rho
_c^*(q_c)/\rho_c(0)$ by extensive Monte Carlo simulations for a
broad range of values for $q_c$. It turns out that for CO$_2$ the
experimental value $Q=4.3 \pm 0.2 D \mathrm{\AA}$ yields

\begin{equation}\label{eq8}
q_c = 0.387,\; \epsilon = 3.491 \times 10^{-21}~J, \quad \sigma =
3.785~\mathrm{\AA} \;,
\end{equation}

while for the case of benzene the value $Q=12 D \mathrm{\AA}$ would imply

\begin{equation}\label{eq9}
q_c = 0.247,\quad \epsilon = 6.910 \times 10^{-21}~J, \quad \sigma = 5.241~\mathrm{\AA} \quad .
\end{equation}

The corresponding results for the vapor-liquid coexistence curves
in the $(T,\rho)$ and $(p,T)$ planes as well as the temperature
dependence of the interfacial tension for both CO$_2$ and
C$_6$H$_6$ were already presented in \onlinecite{36} and shown to give a
rather good agreement with experiments. \cite{56}

Of course, the disregard of the angular dependence of the
quadrupolar part of the interactions is a matter of concern. This
point was investigated by us in \onlinecite{37}, where detailed
comparisons of Monte Carlo results for the full angular-dependent
quadrupole-quadrupole interaction and the isotropic approximation
\{Eqs.~(\ref{eq3})-(\ref{eq6})\} were performed for the case of
CO$_2$. It was shown \cite{37} that the model with LJ + full
quadrupolar interactions (which is still a crude coarse-grained
model, in comparison with all-atom models including partial
charges etc.) does not provide a better account of the
experimental data than the spherically averaged  one.

Another point of concern is the possible sensitivity of the
results of such models to the precise value of $q_c$. Note that
$q$ is proportional to $Q^4$ \{Eq.~(\ref{eq6})\}. Consequently, a small
experimental error in $Q$ is magnified considerably. There may also
be systematic effects since $Q$ is often determined in the dilute
gas phase. Here, we are interested in using densities around
the critical density, and $Q$ could be slightly renormalized
there. 
Packing effects should also be taken into account. 
Indeed, CO$_2$ is not a spherical 
molecule, and at high density a local orientational order could arise.
This packing could enhance some favorable angular correlations that
give rise to a higher effective quadrupolar moment.
One can argue that high temperature perturbative theory
\{see Eq.\ (\ref{eq3})\} may not be very accurate and higher order
terms could be important: in fact our previous investigation
in which a full (angular dependent) quadrupolar interaction was 
considered,\cite{37} proves that this is not the case.
In addition, one may argue that the model of
Eqs.~(\ref{eq3})-(\ref{eq6}) is an effective model, intended for
a good representation of equation of state data, particular for
vapor-liquid equilibria (VLE). Therefore, $q_c$ should be treated as
an effective parameter which can be used to optimize the description of such VLE
data. In this spirit, we have also tried different choices of
$q_c$ and found that a slightly better description of CO$_2$ is obtained

\begin{equation}\label{eq10}
q_c = 0.47\;, \quad \epsilon = 3.349 \cdot 10^{-21}~J, \quad \sigma
= 3.803~\mathrm{\AA}.
\end{equation}

This choice was already included in our previous work.
\cite{36,37} For benzene, a very good agreement with experiments can be achieved for

\begin{equation}\label{eq11}
q_c=0.38\; , \quad \epsilon = 6.472\cdot 10^{-21}~J,\quad \sigma =5.284~\mathrm{\AA}.
\end{equation}

Fig.~\ref{fig6} presents the coexistence curve of benzene in the $\rho - T$
and $T-p$ planes as well as the interfacial tension.
Results based on Eq.~(\ref{eq11}) are compared with results based on
the previous choice \{Eq.~(\ref{eq9})\} \cite{36} and with
experimental data.\cite{56} The description of
the experimental data is clearly remarkable over a wide range of
temperatures. It turns out (see below) that these ``optimized''
choices of parameters \{Eqs.~(\ref{eq10}), (\ref{eq11})\} also yield
a much better description when we consider mixing behavior (e.g.
C$_6$H$_6$ + CH$_4$).

\subsection{Short Alkanes}

In this section we briefly discuss the extension of our
methodology to systems such as propane (C$_3$H$_8$), pentane
(C$_5$H$_{12}$) and hexadecane
(C$_{16}$H$_{34})$. These short alkanes are just
treated as test systems for our methodology and will be used in
Sec. III as components in binary mixtures. Our methodology can be
used, in principle, for any alkanes, provided information on the
vapor-liquid critical point ($T_c^{\exp},\rho_c^{\exp}$) is
available. (Unfortunately, this is not the case for much
longer chains).

As it was already emphasized (fig.~\ref{fig1}) we do not attempt an
all-atom description of alkanes. We also do not use an united atom
model where CH$_2$ (or CH$_3$) groups are described as one
spherical pseudo-atom.\cite{64,65} Such a model requires
torsional and bond angle potentials and is still rather demanding to
simulate. As indicated in fig.~\ref{fig1}, we reduce the
description to a coarse-grained bead-spring model, where a
small number of successive CH$_2$ or CH$_3$ groups are combined
into a single effective monomeric unit. 
For C$_{16}$H$_{34}$ we choose 5 effective units, so
each unit contains about 3 C-C bonds. For pentane and hexane we
choose a dimer (but the effective
LJ parameters $\epsilon$ and $\sigma$ are different, of
course). Such a model is perhaps most questionable in the case
of C$_3$H$_8$, which we treat as a single effective unit (i.e.,
such molecules are treated like almost spherically symmetric
molecules such as methane).

We keep the (truncated and shifted) LJ potential
\{Eqs.~(\ref{eq1}), (\ref{eq2})\} between all pairs of effective
units, bonded and non-bonded ones. In addition we use the well-known FENE potential for the bonded ones  \cite{66}

\begin{equation}\label{eq12}
U_{\textrm{FENE}}(r) = -33.75\epsilon \ln [1-(r/1.5\sigma )^2]
\end{equation}
We note that in Eq.\ (\ref{eq12}) $\epsilon$ and $\sigma$ are
the same parameters as in the LJ potential between 
the monomers. The parameters of the FENE potential 
have been chosen to prevent the
crossing of macromolecules in the course of their motion. 
We note that this choice does not reproduce the characteristic ratio of alkanes accurately.
This means that the FENE potential is fully constrained, and the model
remains a two 
parameter model with parameters chosen to match the
critical temperature and density.

On this coarse-grained level both torsional potentials and
bond-angle potentials between effective beads are ignored.
Hence, it is worthwhile to test whether such crude models are still
able to reproduce the phase diagram and other thermodynamic
properties of the real system correctly. Thus,
figs.~\ref{fig7}-\ref{fig9} show results for the phase diagrams of
several members of the alkane series (including C$_3$H$_8$,
C$_5$H$_{12}$ and C$_{16}$H$_{34}$) in the T-$\rho$ plane, as well
as the corresponding coexistence pressures and interfacial
tensions between the coexisting vapor and liquid phases. The
agreement between the model results and the corresponding
experimental data \cite{56} is remarkable, again, although it is not as
convincing as for methane (which we have included for comparison).
In particular, for C$_5$H$_{12}$  deviations
clearly occur. Table I collects the experimental critical
temperatures, densities, and pressures,\cite{56} as well as our
choices for $\epsilon$ and $\sigma$ for the materials studied, and
the prediction for the critical pressure that results from our
model.

In all cases the critical pressure is predicted with an accuracy of a few
percent, and a glance on Fig.\ \ref{fig8} shows that the slope of the
vapor pressure versus temperature curve is close to the slope derived from experiments, too. For
temperatures away from the critical region (say, 20$\%$ below
$T_c$), deviations 
between experiment and the model predictions become visible, both in the
coexistence curve, coexistence pressure, and interface tension (Fig.\
\ref{fig9}), in particular for propane and pentane. Of course, the accuracy of
the modeling could be enhanced by allowing for additional adjustable parameters 
like in many models in the literature, e.g.\ by introducing a
bond-angle potential, or more interaction sites (see e.g.\
\onlinecite{KT-2004}). Then, quantities such 
as the acentric factor (referring to the shape of the coexistence curve 30$\%$
below $T_c$ \cite{Pitzer_1955})  can presumably be fitted nicely. However, the
simplicity of the coarse-grained model is lost.
Experience with
such somewhat more complicated models shows that these models still require
correction parameters $\xi$ to the LB combining rules that deviate from unity
by about 10$\%$ (see e.g.\ \onlinecite{Vrab2005}). Without these additional
parameters (note that it is not at all straightforward to find optimal values
for these parameters) the gain in accuracy that such models
yield for the description of mixtures is rather modest. Note that an important
motivation for the present work is to develop simple models suitable for the
simulation of polymer solutions (the case of hexadecane in CO$_2$ being just a
prototype case). We are not focusing on pushing the accuracy of modeling of
pure short alkanes to its limit.

\section{PHASE BEHAVIOR OF SELECTED BINARY MIXTURES}\label{secmixture}

Extending our treatment to binary systems (A,B) one wishes to
describe the interactions between unlike particles by a potential
of the same functional form as it is used for the interactions
between particles of the same type, i.e. the Lennard-Jones
potential in our case. The simplest choice, most often used in the
literature, is the Lorentz-Berthelot combining rule \cite{52}

\begin{equation}\label{eq13}
\sigma_{AB} = (\sigma_{AA} + \sigma_{BB})/2,\quad \epsilon _{AB} =
\sqrt{\epsilon_{AA}\epsilon_{BB}}
\end{equation}

As is well-known, there is really no convincing derivation of
Eq.~(\ref{eq13}), so there is no reason to believe that Eq.~(\ref{eq13})
is exact. At best it is a practically useful approximation. As a
matter of fact, several alternatives to Eq.~(\ref{eq13}) have been
proposed in the literature.\cite{45,52,67,68,69} Although it has
been demonstrated that there are some cases where some of these
alternative combining rules work better, in general none of these
alternative combining rules has a really clear advantage.\cite{41}
Since we wish to explore a very simple and general approach, we do
not implement any alternatives to the simple Lorentz-Berthelot
rule in our paper, even when one has to pay the price of
sacrificing a small improvement in the accuracy of our modeling.
We also note that the Lorentz-Berthelot rule works very well
for the prediction of virial coefficients for the mixture of Argon
plus CO$_2$, a mixture of an apolar and a quadrupolar fluid.\cite{69´}
We want to stress that proceeding in such a way no experimental input
from the mixture phase diagram is required for
testing a full predictive model for the mixture. This also holds for 
the TPT1 computations  which require only
$\epsilon$ and $\sigma$ that can be obtained  using Monte Carlo results
of the pure component critical line.\cite{36} 
Coexistence densities and pressure have been computed as in 
pure component systems.\cite{24} On the other hand, the computation
of the critical points is more complicated. Indeed,
 in a binary mixtures close to criticality the
proper identification of the order parameter is a subtle problem.
\cite{70} In principle, complete scaling \cite{71,72,73} in
the case of binary mixtures implies that three scaling fields
occur, which are linear combinations of four independent intensive
variables: the deviations of two chemical potentials, temperature,
and pressure from their values at the critical point.
Consequently, the order parameter density becomes a function of
the appropriate conjugate variable, and the relevant physical
densities (particle number densities, entropy density) become
nonlinear functions of the proper scaling fields.\cite{70}
Since this formalism is somewhat cumbersome for the case of 
compressible binary fluid mixtures,
we simplify the problem by applying ``field-mixing''-procedures
analogous to the method of Wilding \cite{19,20,21} which is
rather successful for most one-component fluids.
Details on this procedure are reported in the appendix
\ref{appendix}, presenting the analysis done for a critical point
of the Krypton Xenon mixture. In order to estimate systematic 
errors of this procedure, in appendix \ref{appendix} we also present
 results with a full finite size analysis with cumulants
crossing \cite{16} for a highly asymmetric mixture
like carbon dioxide in hexadecane.

\subsection{Mixtures of Small Apolar Molecules}

As a first example of apolar mixture we present results for 
kripton plus xenon.
As it has been discussed in Sec.~II, the noble gases already exhibit rather
large deviations between the experimental data and the model
calculations based on the Lennard-Jones potential. Thus, it is
interesting to see whether these problems get even worse when
mixtures are considered.
The resulting critical line in the ($p,T$) plane
is  shown in fig.~\ref{fig4} for both choices of $\epsilon$
and $\sigma$ as discussed in Sec.~II A. 
If we fit $\rho_c$ and $T_c$ for the pure systems, the
predicted critical points for the mixture 
deviate from the experimental curve about as much as for the pure
systems. If we adjust $\epsilon, \sigma$ such that $p_c,T_c$ is
reproduced, the data \cite{59'} for the two mixed systems are
almost perfectly reproduced. 
The variation
of the critical concentration with temperature is also rather well
reproduced (fig.~\ref{fig11}) by both models where $\rho _c$ and $T_c$ 
or $p_c$ and $T_c$ are fitted to experimental values.

As a second case we consider now methane in butane.
In Secs.~II B, C we showed that the simple LJ model gives
a fairly accurate account of the equation of state of both CH$_4$
and C$_3$H$_8$. Therefore, it is natural to consider a mixture of those two
molecules as a next step. Of course, a comprehensive study of the
phase behavior of such mixtures in the space of all three
variables $(T,p,x)$ is a nontrivial effort. Therefore we limit
ourselves to consider only isothermal slices through the phase
diagram, following a standard practice in the literature.
\cite{41,43} As an example, fig.~\ref{fig12} shows two such
slices at $T=327K$ (a) and $T=277 K $ (b), and compares
experimental data \cite{74} with selected Monte Carlo data and results
from our implementation of the TPT1-MSA (which is described in 
appendix B of \cite{36}). 
We emphasize that the various parameters characterizing
the interactions among the various molecules are those obtained
from Monte Carlo simulations of the pure materials (Sec.\ II),
together with the Lorentz-Berthelot rule. 
These parameters also serve as input for
TPT1-MSA: there are no additional parameters that 
enter the latter approach. 
Thus we present comparisons between experiments, 
simulations and theory in which no adjustable parameters for the 
mixture have been used.

It should be noted that both chosen temperatures in fig.\ 
\ref{fig12} fall
below the critical temperature of C$_3$H$_8$ but exceed the
critical temperature of CH$_4$. Therefore, the characteristic
bubble-shaped liquid-vapor coexistence curve results, starting out
at the ordinate axis at the vapor-liquid coexistence point of pure
C$_3$H$_8$, but not extending to CH$_4$ concentrations close to
$x=1$. The critical point occurs at the maximum of this closed
loop. (The liquid phase is located on the upper part of the loop to the
left of the critical point, the remaining part of the curve
describes the vapor). For $T=327K $ and $x \leq
0.35$ both experiment, TPT1 and Monte Carlo agree nicely. For
larger $x$, however, a systematic discrepancy between Monte Carlo data
and experiment shows up. The TPT1-MSA approximation overestimates
the critical pressure substantially. This problem already
occurs in the pure systems, as is well-known, and is an inevitable
consequence of simple mean-field-like approximations.
\cite{36,53,54} Fig.~\ref{fig5} shows that the critical
temperature and pressure of pure CH$_4$ are both overestimated.
The same holds for pure C$_3$H$_8$, and the whole line of
critical points $T_c(x)$ that connects $T_c(0)$ and $T_c(1)$ when
we would project them into the $(p,T)$ plane as we did for the Kr-Xe
mixture (fig.~\ref{fig4}). As in the latter case, the mixture of
CH$_4$ and C$_3$H$_8$ has a simple ``type I'' phase diagram in the
classification scheme of fluid binary mixtures \cite{38,39,40}
(type 1$^P$ in the modern classification \cite{D_98}). As
a consequence, we expect that TPT1-MSA predicts too large
vapor-liquid coexistence loops in the $(p,x)$ plane at all
temperatures that are supercritical for CH$_4$ but subcritical for
C$_3$H$_8$.

A more disturbing discrepancy seems to occur between the data
\cite{74} and the theoretical results at the lower temperature
$(T=277K)$, where at small $x$ the vapor pressure at coexistence
falls slightly but systematically below the experimental data. For
molar concentrations well below criticality, Monte Carlo results
and TPT1-MSA agree very well, and our numerical
procedures are accurate for our model. Hence, assuming that
the experimental data are accurate enough so that the discrepancy
is meaningful, this result indicates that some limitations of our
model become apparent. This is not really a surprise, of course,
because in the data for pure propane at this temperature
discrepancies of the order of a few percent do occur as well
(figs.~\ref{fig7}-\ref{fig9}).

As a third case we now consider the mixture of CH$_4$, and
C$_5$H$_{12}$, because for pentane slightly larger deviations
between the predicted and observed coexistence vapor pressure do
occur over a much broader temperature range (fig.~\ref{fig8}).
Indeed, the corresponding isothermal slices through the phase
diagram of that mixture (fig.~\ref{fig13}), which still is a
type-I phase diagram, show that slight but systematic
discrepancies are now seen at the higher temperature as well. At
the low temperature, the phase diagram can only be reproduced in a
rather qualitative manner. Note, however, that $T=237K$ is less
than 50\% of the critical temperature of pentane, where the
effective interactions of pentane were adjusted: of course, the
coarse-grained modelling used in our work should not be pushed to
too low temperatures. Keeping this limitation in mind, we conclude
that a rather satisfactory description of mixing behavior of these
systems is in fact reached by our models. Hoping for
perfect agreement would have been premature, in view of the
simplicity of our models. But the phase diagram predictions should
allow a useful first orientation at temperatures not too far below
of the higher critical temperature of the components in such a binary
mixture.

\subsection{Mixtures of small molecules, one of which has a quadrupole moment}

We begin with a mixture of CH$_4$ and CO$_2$,
because for both pure molecules a particularly accurate
description of the equation of state was
obtained (see Sec.\ II). Again we note that the CH$_4$ + CO$_2$
system belongs to the category of ``type I'' phase diagram in the
classification scheme of Scott and van Konynenburg \cite{38,39,40}
(1$^P$ in the modern classification \cite{D_98})
and the temperature regime of interest for our modeling is the
regime in between the critical temperatures of the two
constituents of this mixture. Note that Eq.~(\ref{eq13}) only applies to
the LJ part of the interactions of CO$_2$, since CH$_4$ has
no quadrupole moment.

In fig.~\ref{fig14} we present isothermal slices through the
phase diagram in the space of variables $(T,p,x)$. If one uses TPT1-MSA
the model for CO$_2$ based on Eq.~(\ref{eq10}) can describe the
mixing behavior with CH$_4$ very accurately at molar
concentrations $x$ of CH$_4$ and pressures that are not close to
criticality. As emphasized above,
mean-field theories such as TPT1-MSA are not expected to be
accurate near critical points. Hence, the discrepancy that
TPT1-MSA predicts a too large loop inside of which two-phase
coexistence occurs, is inevitable and expected. But for the model
Eq.~(\ref{eq10}) the part of the loop at not too large $x$ is
significantly more accurate (full curves) than a simple LJ model
for CO$_2$ would be (broken curves). As expected, at low
temperatures (such as $T=230K$) the quadrupolar model for CO$_2$
\{Eq.~(\ref{eq8})\} also starts to show slight but systematic deviations
from the experiment at the vapor branch of the vapor-liquid
coexistence curve. This is similar to our finding for the apolar
mixtures (Sec.~III B).

In order to verify that the good agreement between experiment and
theory for the quadrupolar model of CO$_2$ in
the CH$_4$+ CO$_2$ mixture is not just fortuitous, we
show in fig.~\ref{fig15} corresponding results for the mixture of
benzene (C$_6$H$_6$) and methane (CH$_4$). This is a more
stringent test, since the critical temperatures of the two
constituents are rather far apart from each other (cf.\
figs.~\ref{fig5}, \ref{fig6}). Nevertheless, the conclusions are
the same as in the case of CH$_4$+CO$_2$: using interaction
parameters that were optimized for the pure systems, namely those
of Eq.~(\ref{eq11}) in the case of C$_6$H$_6$, and adjusting them to
Monte Carlo  results as described in Sec.~II, we can proceed to
the description of the mixture data \cite{78} and estimate the
missing mixed interaction parameters from the Lorentz-Berthelot
rule, Eq.~(\ref{eq13}). The use of these interaction parameters in a
simple and fast analytical theory for the EOS such as TPT1-MSA
then provides a satisfactory description of the phase behavior of
the mixture, apart from the vicinity of critical points (this
drawback can be rectified by carrying out MC work for the mixture
as well, of course) and for not too low temperatures. (For
temperatures of the order of 50\% of the critical temperature
$T_c$ of the constituent with the higher $T_c$ systematic
deviations start to appear rather generally.)

The last example of this section deals with a slightly more
complicated case, namely the CO$_2$ + C$_5$H$_{12}$ system
(fig.~\ref{fig16}): while CO$_2$ is still represented as a point
particle with a quadrupole moment, as in the previous examples,
the other partner of this mixture (C$_5$H$_{12}$) should not be
coarse-grained into a point particle any more, but rather needs to
be represented as a dimer (i.e., a dumbbell-like effective
molecule). In this case the TPT1-MSA theory predicts
unmixing over a far too large range of molar CO$_2$
concentrations, and the improvement provided by the inclusion
of the quadrupolar moment at small $x$ is only qualitative, but
not quantitative. On the other hand, the Monte Carlo results for
this model are in rather good agreement with the corresponding
experimental data.\cite{80m} Since MC and TPT1-MSA are using
precisely the same interaction parameters, we conclude that
for this particular case TPT1-MSA is somewhat inaccurate for the vapor branch
of the mixture, far away from criticality. A related discrepancy
was already noted for the CH$_4$ + C$_5$H$_{12}$ system at $T =
378K$ (fig.~\ref{fig13}a). Perhaps this indicates that TPT1-MSA
does not capture the statistical mechanics of flexible dimers well
enough.

\subsection{Polymer solutions: The CO$_2$+C$_{16}$H$_{34}$
system revisited}

Virnau et al.~\cite{44,44a,44b} already attempted to model this system,
describing CO$_2$ as a point particle with no quadrupole
moment. They found that using the Lorentz-Berthelot rule
\{Eq.~(\ref{eq13})\} the phase diagram predicted by the model
belongs to type I, while experiments suggest
\cite{78,79} that this system belongs to the type III class
(1$^C$1$^Z$, according to \onlinecite{D_98}, where 1$^C$ means that 
the critical line emanating from the pure component critical point of the
hexadecane goes to high pressure regions without joining the solvent critical 
point like in diagrams starting with 1$^P$).
Virnau et al. \cite{44} proposed that one can improve the
description by using an empirical factor $\xi$ to modify
Eq.~(\ref{eq13}), assuming that $\epsilon_{AB} = \xi
\sqrt{\epsilon_{AA}\epsilon_{BB}}$ instead of
$\epsilon_{AB}=\sqrt{\epsilon_{AA}\epsilon_{BB}}$. 
In the literature, the value of $\xi$ depends on the specific 
mixture and typically is written in the form  $\xi_{AB}=1-k_{AB}$, 
with $k_{AB}\ge 0$. Of course,
there is not really a theoretical justification for doing so, and
$\xi$ simply plays the role of a fitting parameter. By trial and
error it was found that $\xi = 0.886$ provides a description
compatible with the experimental data.

In the present subsection of our paper, we show that the main
source of the problems encountered in \onlinecite{44} was the neglect of
the quadrupole moment. Thus, we have repeated the study of the
CO$_2$+C$_{16}$H$_{34}$ system, insisting on the Lorentz-Berthelot
rule, Eq.~(\ref{eq13}), but using Eq.~(\ref{eq10}) as an improved
model for CO$_2$, as in the previous subsection. Again, the
Lorentz-Berthelot rule is only applied to the Lennard-Jones part of the
interactions, since C$_{16}$H$_{34}$ does not have a
quadrupole moment.

Following the strategy of the previous subsection, we have
computed an isothermal slice through the phase diagram at $T=486
K$, where data from the previous simulation \cite{44} were
available both for $\xi=1$ and for $\xi = 0.886$. Indeed it is
found that the data of the present model ($\xi=1$, but optimized
quadrupolar interaction $q_c=0.47$ for pure CO$_2$) are well
compatible with the experimental data \cite{80} and almost fall on
top of the results of the previous calculation with $q_c=0$ and
$\xi=0.886$.\cite{44}

Of course, we have already seen in the previous subsections, that
often a very good agreement between our description based on a
very simplified model occurs at high enough temperatures. In order
to test, to what extent this problem arises for the present
system, we have followed the strategy of \onlinecite{44} to compute the
full critical line $T_c(x),p_c(x)$ for the full range of molar
concentrations $x$ of CO$_2$. Fig.~\ref{fig18} shows the resulting
projection into the $p^*,T^*$ plane. (Here $p,T$ are given in LJ
units, with the LJ parameters of the effective monomers used to
rescale the variables). One sees that the simulations with nonzero
quadrupole moment included in the figure are close to those for
$\xi=0.9$, $q_c=0$, for $T^*\leq 1.3$. As a consequence, the model
that we have developed for CO$_2$, Eq.~(\ref{eq10}), is still 
not able to yield the correct phase diagram topology. 
(For $\xi=0.9$, transition type IV was observed in Ref.~\onlinecite{44}, as opposed to type III which was observed experimentally.)
For $T^* <0.8$ the model does not
yet describe the properties of hexadecane + carbon
dioxide mixtures accurately, although for $T^*\geq 1.3 \; (T \geq 545 K)$ the
properties of the system are predicted rather satisfactorily. Of
course, this result is not unexpected. 
For $T\leq
0.5T_c^{hex} \approx 360 K$ the model based on fitting the
critical parameters of hexadecane to fix its interaction
parameters starts to become inaccurate. 
On the other hand, the proper prediction of the phase diagram type is 
a very stringent test. Indeed, variation of the interaction parameters
by a few percent could drastically change the type of 
the phase diagram. \cite{44}

\section{Conclusions}
In this paper, we have studied the phase diagrams of a variety
of fluid binary mixtures, with particular emphasis on mixtures
of alkanes in supercritical carbon dioxide and benzene. In order to
better understand the performance of our modeling for these systems,
we have also investigated mixtures with apolar solvents including
noble gases and methane. 
We have investigated the accuracy of the use of the Lorentz-Berthelot
rules for describing the mixing behavior, based on interaction parameters
for the pure systems that are tuned such that the critical point
(critical temperature, critical density or pressure) of the pure systems 
are well reproduced.
Using a simple Lennard-Jones model for interaction parameters of
pure apolar fluids, Monte Carlo calculations in the grand-canonical
ensemble, analyzed by appropriate finite size scaling methods, readily yield
the desired accuracy for this procedure. 
For the polar molecules we use a spherically averaged point-like quadrupolar
interaction, \cite{62,63,36} which was shown to produce very good
phase diagrams, \cite{36} also if compared to more realistic atomistic 
models. Our model takes as experimental input the critical temperatures
and densities of the pure components (like in previous coarse grained  schemes
\cite{44}) plus the experimental quadrupolar moments.
For pure CO$_2$ and C$_6$H$_6$ this choice leads to a 
significant improvement in comparison with a simple LJ model
without explicitly accounting for the polar interactions. In 
Ref.\ \onlinecite{36} and Fig.\ \ref{fig6},  as a second option, we
have treated the quadrupole moment as an effective parameter
in an attempt to optimize agreement with experiments. 
We tune this parameter such that the liquid branch of the 
vapor-liquid coexistence curve of pure CO$_2$ or pure C$_6$H$_6$
is optimally represented. In the case of benzene, for which 
the optimization procedure seems to work very well, 
the agreement with the coexistence pressure is also improved. (This is not
the case of CO$_2$ which is however better described than benezene
if the experimental values for the quadrupole moments are used.) 
The physical reason for this requirement to work with an effective
quadrupole moment is presumably that actual molecules are
not point-like particles, of course: CO$_2$ is a rather elongated molecule,
while C$_6$H$_6$ is disk-like. So packing effects should occur, i.e.\
local orientational correlations, which are underestimated by the 
quadrupolar interaction. 
Thus, it is gratifying to note that  
a remarkable improvement of accuracy in the prediction of
the phase behavior of mixtures is achieved if this effective quadrupole moment is used.

These energy parameters, which we fixed from the description of
the pure systems, together with the Lorentz-Berthelot rules, allow us 
to predict phase diagrams of mixtures, with no ambiguity whatsoever,
since no further adjustable parameters occur. Two methods of prediction
are used: (i) Monte Carlo simulations (ii) TPT1-MSA calculations. The Monte Carlo approach
has the substantial advantage that it is also accurate near critical 
points of the mixture. In principle, we obtain the exact statistical 
mechanics of the model system. Any discrepancy between experiment and prediction
is entirely due to a shortcoming of the (simplified) model. 
The TPT1-MSA approach has the merit that relatively little computational
effort is necessary to implement it. However, it clearly involves
various approximations and hence the interpretation of discrepancies 
between TPT1-MSA and experiment is not so clear - part of them being
due to inadequacies of the model, part of them stem from inaccurate 
approximations. For instance, TPT1-MSA, like all mean-field theories,
overestimates the critical temperature and pressure, so the 
isothermal slices through the phase diagram of the mixture always
involve two-phase regions which are too large. 

We note that fluids like CO$_2$ have an important application as 
supercritical solvents. If one aims at describing the behavior in 
the critical region of the pure solvent and of the mixtures 
correctly, Monte Carlo methods have a clear advantage. 
Now, one could try to readjust parameters in the TPT1-MSA approach 
to improve agreement with experiment (like done for instance in 
\onlinecite{McD_2002}, where $\epsilon$ and $\sigma$ for the EOS have
been rescaled  in comparison 
to the model used in MC simulations in order to properly reproduce 
the critical points of the pure compounds), but this would be just an 
attempt to provide a partial cancellation of errors, and in other regions 
of the phase diagram the description would necessarily get worse. 
\cite{McD_2002} Since we feel that relatively little  physical insight 
is gained by such fitting procedures, they have not been implemented in our
 paper.
Our overall conclusion is that in the framework of the modeling
as defined above the Lorentz-Berthelot rules work very well, in
the sense that an ad-hoc change of mixed binary interactions by
at most a few percent (typically one or two percent) would lead
to almost perfect agreement with experiment.
As a piece of evidence for this claim, we note that in the study
of the CO$_2$+C$_{16}$H$_{34}$ system by Virnau et al., \cite{44}
where the CO$_2$ molecule was modeled as a point particle with 
LJ interactions with no account of the quadrupole moment, a 
correction factor $\xi=0.886$ to the Lorentz-Berthelot rule 
was required to produce good agreement with experiment.
However, the present model (with a quadrupolar interaction and
no correction factor) yields results that are almost identical to 
those of Virnau et al.\ \cite{44} when $\xi=0.900$ is chosen.
As a consequence, we conclude that in the present model a
correction factor $\xi\approx0.985$ would suffice to reproduce the results.
Noting that  the Lorentz-Berthelot rule assumes that the
mixed correlation functions in the fluid described 
atomistically behave in the same way as in the coarse-grained
descriptions, deviations of such a fitting parameter $\xi$
from unity in the range from 1 to 2\% are no surprise at all.

Thus we feel that the present level of accuracy cannot easily
be improved in the framework of our model. As it has been emphasized
already in the introduction, many more complicated models for fluids
are discussed in the literature (see \onlinecite{28,29,30,31,32,33,81}).
Optimizing parameters in those models such that the critical properties
of the pure systems and their vapor-liquid coexistence curves are
very well reproduced, might be an alternative starting point 
to test the validity of combining rules.
\cite{Vrab2001,Stoll2003,Vrab2005}  However, even for our very 
simple model the Monte Carlo runs require substantial computer
resources, and hence we have not attempted to generalize our 
approach to other models. 
The strength of our method, if compared to more detailed
models with a lot of parameters the optimization of which would require 
massive computation, is its generality. It is possible to
have the potentials for a given mixture, without any extra computational
efforts from the results of the pure components. \cite{36} 
It is also important to mention that the present work 
validates the use of spherical averaged quadrupolar potential. \cite{62,63,36}

Finally, it is important to observe that our way to coarse grain solvent
molecules into single beads has the advantage, with respect to atomistic models
like multi center Lennard Jones, to be accessible to advanced equation of
state machineries. In this paper we have shown how, with rather small efforts,
significant results can be obtained. We are also aware of the fact that
several improvements could be done (e.g.\ using some integral equation scheme
which should improve the MSA solution near the critical point). However, most
of the advanced methods in equation of state modeling, apply only for
reference systems that are mixture of monomers (i.e.\ beads with point-like
interactions), the associating part being taken into account by TPT1. On the
other hand TPT1 gives a reasonable description only if in the ``associated
molecule'' diameters of the beads do not overlap. This is of course the case
in our CG model for alkanes in which the experimental distance between three
carbon units (d=4.59 \AA) is bigger than the typical $\sigma$ used
($\sigma\approx$4 \AA). 
(This is another reason the FENE potential uses the same simulation parameters
of the LJ interaction.) The condition d$>\sigma$ guarantees that the reference
system (a mixture of monomers) is a good starting point for a perturbation
theory. On the other hand if d$<\sigma$ association is too strong and cannot
be properly taken into account by TPT1 (i.e.\ the monomer reference system is
not the adequate starting point for a perturbation expansion). In models which
describe simple solvent molecules with several interacting points (see e.g.\
Tab.\ 1 of \onlinecite{Vrab2001} for typical parameters of two center LJ) we have
d$<\sigma$: this implies that these models cannot be investigated with
associating theories. In conclusion, our modeling approach might enable
the application of modern EOS which is another important motivation of our work.

We do feel that the approach 
based on the present models is able to make nontrivial and 
practically useful predictions for a large class of
systems. Hopefully more experimental data on mixtures will become available to
allow for more stringent tests.
\\
\\
{\bf ACKNOWLEDGEMENTS}
CPU times was provided by the NIC J\"ulich and the  ZDV Mainz.
We would like to thank  M.\ Oettel (University of Mainz), F.\ Heilmann and H.\ Weiss (BASF AG, Ludwigshafen) for fruitful discussions.
BMM would also like to acknowledge BASF AG (Ludwigshafen) 
for financial support. 
LGM wishes to acknowledge support from Ministerio de Educacion y
Ciencia 
(project FIS2007-66079-C02-01) and Comunidad Autonoma de Madrid
(project MOSSNOHO-S0505/ESP/0299).

\appendix
\section{Determination of the critical point}\label{appendix}

In this appendix we describe how the critical points for the mixtures
(e.g.\ figs.\ \ref{fig4} \ref{fig18}) have been obtained. 
As mentioned in the sec.\ \ref{secmixture} in asymmetric binary mixtures
the determination of the order parameter is difficult  \cite{70}
and a complete scaling \cite{71,72,73} requires a lot of
work which has to be repeated for every critical point.
For this reason in the present work 
the method of Wilding \cite{19,20,21} has been used. 
Taking as an example the kripton xenon mixture (but the
same procedure has been done for all the other mixtures),
we take the order parameter $M$ as a linear
combination of the particle numbers of Kr atoms ($N_k$) and of
Xe atoms ($N_x$) and of the total potential energy $E_{tot}$
\begin{equation}\label{eq14}
M=N_x+x_1N_k+x_2E_{tot}\; ,
\end{equation}
where the parameters $x_1,x_2$ are determined by the following
iterative procedure. First, the chemical potential $\mu_x^*$ (in
LJ units) is tuned to get vapor-liquid phase coexistence (i.e.,
the distribution $P(M)$ satisfies the equal area rule). Normally,
due to the lacking of the particle--hole symmetry,
the two peaks will not be symmetric such as the two peaks of the
universal Ising model distribution at criticality. Thus, as a second step
the other chemical potential is also tuned (and the
step 1 is repeated), so that one gets somewhat closer to the critical
point of the system. Still, the universal shape of the
distribution is not yet obtained. The third step consists in a
variation of $x_1$ (and repeating steps 1 and 2) such that the two
peaks of the distribution become as similar to each other as
possible. The fourth step amounts to a variation of $x_2$
(again repeating steps 1 and 2). In this way (for the investigated
cases) it is possible to
obtain a final histogram of the order parameter $M$ that reproduces
the universal Ising shape at criticality almost exactly
(fig.~\ref{fig10}).

On the other hand the previous approach is not totally 
correct because it neglects finite size corrections for
the critical parameters but simply ``supposes'' that
the simulation box is large enough. In order to elucidate this
point in this appendix we also report the full finite size
analysis \cite{16} with crossing cumulants for several simulation 
boxes. We do this investigation for the polymer solution studied
in this paper (CO$_2$+C$_{16}$H$_{34}$) which should be more 
sensible to mixing effects being a highly asymmetric mixture
in which
the coupling between two order parameters (total density and
relative concentration, respectively) may be more of a problem
rather than for some noble gas or small molecule mixtures. 
In fig.\ \ref{fig19} we report our finite size analysis for 
the critical point at T=486 K.
From extensive $\mu _s \mu _p VT$
simulations ($\mu_s$ being the chemical potential of the solvent,
CO$_2$ in this case, and $\mu _p$ the chemical potential of the
polymer C$_{16}$H$_{34}$) histograms have been generated for the
total energy $E_{tot}$, the number of solvent particles $N_s$ and
the number of polymers, $N_p$. The probability distribution for
the order parameter $M=N_p+x_1 N_s+x_2 E_{tot}$ is computed, using
$x_1=0.08$ as an initial guess (it turns out that the final
results depend on the parameter $x_2$ so weakly, that one may
choose $x_2=0$ here; on the other hand the choice $x_1=0.08$ was suggested
comparing $P(M)$ with the universal Ising curve, similarly to what 
has been done above for the Krypton-Xenon mixture). 
The simulation box linear dimensions were
 $L=9 \sigma _p$,  $L=11.3 \sigma _p$ and $L=13.5 \sigma _p$.
For a fixed $x_1$ and $\mu_s$, $\mu_p$ is always fixed
so that $P(M)$ satisfies the equal area rule.
Then, we compute second and fourth order cumulants $B_2=\langle 
{\mathcal{O}}^2 \rangle /\langle |{\mathcal{O}}|\rangle^2$,
 $B_4=\langle 
{\mathcal{O}}^4 \rangle /\langle {\mathcal{O}}^2\rangle^2$
(where $\mathcal{O} = M-\langle M \rangle$)
  as a function of $\mu_s$, for
different $L$ (fig.~\ref{fig19}a). It is seen that rather
well-defined intersection points of the curves $B_2$ and $B_4$ vs.
$\mu_s$ for different choices of $L$ do in fact occur at $\mu_s=
-2.058 \pm 0.001$. 
Using a simple comparison with the universal critical
curve (which is the method we have used for all the critical lines 
of fig.\ \ref{fig18}) we previously obtained a value $\mu_s=-2.06$,
which is very close to the value determined above
(in particular within the 0.5\% which is our general estimate
of errorbars).
The intersection for $B_2$ occurs close to the
theoretical value \cite{21} $B^*_2= 1.22382$, while the intersection
point for the curves $B_4$ is somewhat too low. This may indicate
that the choice of the mixing parameter $x_1$ is not optimal.
However, varying $B_4$ as a function of $x_1$, at fixed choices 
of $\mu _s$ we observe that $B_4$
has a shallow minimum near the chosen value $x_1=0.08$ (see inset of 
fig.~\ref{fig19}a). This could justify in some sense the apriori chosen 
value for $x_1$ and gives an estimate of the systematic error related
to the choice of $x_1$, which is very small.
(We notice that variation of $B_4$ with $x_1$ is not strong enough that one
could tune the parameters such that the intersection occurs
strictly at the theoretical value, and with a distribution function
$P(M)$ which is far from looking like the Ising curve). 
We conclude that all the sizes
from $L=9 \sigma _p$ up to $L= 13.5 \sigma _p$ are not yet in the
asymptotic region of finite size scaling, and so various
corrections to finite size scaling occur which could only be
disentangled if a much wider range of $L$ were at our disposal. In
view of such possible systematic errors, we have allowed an error
bar for the critical values of $\mu_s$ and $\mu_p$ of 3 parts in a
thousand, three times as large as one would conclude from a naive
analysis of fig.~\ref{fig19}a. In any case, the uncertainties
resulting from these finite size analysis are much less than the
deviation between our model predictions and the experimental data.

\newpage

\begin{center}
\begin{table}[h]
\caption{We report all the simulation parameters used in the present 
work and
the critical parameters of the pure components 
that have been used. In brackets 
we also report the errors (if $n$ digits are reported, 
the error applies to the same last digits). 
The errors for $\epsilon$ and $\sigma$ have been estimated using the 
experimental errors for the critical points \cite{56}
and an error of 0.5\% in the simulation critical points as 
estimated previously.\cite{36}  As a consequence we note for instance that
in the three models used for C$_6$H$_6$ the three values for $\sigma$ are 
almost compatible with our error bar.
It is important to observe that in this discussion we have disregarded 
the huge error in the quadrupolar moment Q (and as a consequence in $q$).
 We refer to the text for a discussion of this point.
}
\end{table}
\label{table}
\end{center}

\begin{figure}[h]\caption{\label{fig1}(Color online) Illustration of the
coarse-graining procedure: in the case of hexadecane, three
successive C-C bonds are integrated into one bead (dotted circle).
The oligomer, containing 50 atoms (or 16 ``united atoms'', CH$_3$
or CH$_2$, respectively) is thus reduced to an effective chain of
5 beads. Neighboring beads along a chain interact with a
combination of Lennard Jones (LJ) and finitely extensible
nonlinear elastic (FENE) potentials. Non-bonded beads only interact
with a single LJ potential. Carbon dioxide is represented by
a point particle, which carries a quadrupole moment.}
\end{figure}


\begin{figure}[h]\caption{\label{fig3}(Color online) Coexistence curve for Kr
(upper curve) and Xe (lower curve) in the temperature
density-plane (a), and interface tension plotted vs. temperature
(b). Broken curves indicate the experimental data,\cite{56}
asterisks MC results where $\rho_c^{\exp}$ was used to adjust
$\sigma$, while circles show data where $p_c^{\exp}$ was used to
adjust $\sigma $ (cf.~ text).}
\end{figure}

\begin{figure}[h]\caption{\label{fig4}
(Color online) Coexistence pressures for pure
 Krypton (right) and Xenon (left) plotted vs.~temperature. The
 broken line shows the experimental data for the pure noble gases,
 while the full curve is the projection of the critical line of
 the binary mixture (to be discussed in Sec.~III.A).\cite{59}
 The symbols denote our simulation data (the notation is the same
 as in fig.~\ref{fig3}).}
\end{figure}

\begin{figure}[h]\caption{\label{fig5}(Color online) Coexistence curve for CH$_4$
in the temperature-density plane (a), vapor pressure at
coexistence (b) and surface tension plotted vs. temperature (c).
Broken curves show experimental data (Ref.~\onlinecite{56}), crosses the Monte
Carlo simulation results, while the full lines show the MSA predictions.}
\end{figure}

\begin{figure}[h]\caption{\label{fig6}(Color online) Coexistence 
curve describing
vapor-liquid equilibrium for benzene (C$_6$H$_6$) in the
temperature density plane (a), temperature dependence of the vapor
pressure (b) and the interfacial tensions (c) at phase
coexistence. The full curve is the result of fitting $T_c^{\exp}, \;
\rho_c^{\exp}$ \cite{56} to a simple LJ model without taking into
account any contribution from quadrupolar interactions, while
triangles are Monte Carlo results using Eq.~(\ref{eq9}) and
crosses Eq.~(\ref{eq11}), respectively. Broken curves are the
experimental data.\cite{56} }
\end{figure}

\begin{figure}[h]\caption{\label{fig7}(Color online) Coexistence densities 
for the
alkanes studied in the present paper. The following substances are
reported (from below): Methane (CH$_4$), Propane (C$_3$H$_8$),
Pentane (C$_5$H$_{12}$)  and Hexadecane
(C$_{16}$H$_{34}$). Curves are experimental data,\cite{56}  while
the open circles are our simulation results.}
\end{figure}

\begin{figure}[h]\caption{\label{fig8}(Color online) Coexistence 
pressures plotted
vs. temperature, for the alkanes studied in the present paper,
namely CH$_4$, C$_3$H$_8$, C$_5$H$_{12}$ and
C$_{16}$H$_{34}$ (from left to right). Curves are experimental
data,\cite{56} dots show our simulation results.}
\end{figure}

\begin{figure}[h]\caption{\label{fig9}(Color online) Interface tensions
    plotted vs.\ 
temperature, for the alkanes studied in the present paper, namely
CH$_4$, C$_3$H$_8$, C$_5$H$_{12}$ and
C$_{16}$H$_{34}$ (from left to right). Curves are experimental
data,\cite{56} dots show our simulation results.}
\end{figure}

\begin{figure}[h]\caption{\label{fig11}(Color online) 
(a) Krypton concentration  at
criticality plotted vs. temperature; 
the curve shows the experimental data,\cite{59'}
while symbols denote our simulation data (the notation
is the same as in fig.\ \ref{fig3}).
}
\end{figure}

\begin{figure}[h]\caption{\label{fig12}(Color online) Isothermal 
slice through the
phase diagram of the mixture of CH$_4$ and C$_3$H$_8$ at $T=327K$
(a) and $T=277 K$ (b), using the molar fraction $x$ of CH$_4$ as
abscissa variable and pressure $p$ as the ordinate variable. Dots 
are experimental data \cite{74} broken curves result from
our TPT1-MSA approximation, and symbols denote Monte Carlo data
(see text). Triangles are MC results for the critical points.}
\end{figure}

\begin{figure}[h]\caption{\label{fig13}(Color online) 
Isothermal slice through the
phase diagram of the mixture of CH$_4$ and C$_5$H$_{12}$ at
$T=378K$ (a) and $T=237K$ (b). Triangles denote experimental 
data,\cite{75,76} broken curve denotes TPT1-MSA, and asterisks denote
Monte Carlo results (which were only taken for $T=378K$).
Triangles are MC results for the critical points.}
\end{figure}

\begin{figure}[h]\caption{\label{fig14}(Color online) 
Isothermal slices through the
phase diagram of the CH$_4$+CO$_2$ system at $T=270K$ (a), $T=250
K$ (b) and $T=230 K$ (c). Dots represent experimental data
\cite{74} while the broken curves are the results of TPT1-MSA when
CO$_2$ is represented as a point particle with no quadrupole
moment $(q_c =0)$. The full curves are results of TPT1-MSA with
the spherically averaged quadrupolar interaction using the
parameters of Eq.~(\ref{eq8}) $(q_c=0.387)$.
The triangle shows the MC result for the critical point.}
\end{figure}

\begin{figure}[h]\caption{\label{fig15}(Color online) Isothermal 
slices through the
phase diagram of the CH$_4$ + C$_6$H$_6$ systems at $T=501.15K$
(a), $T=461.85K$ (b) and $T=421.05K$ (c). Full dots show
experimental data \cite{77} curves are calculations based on
TPT1-MSA: broken curves denote the simple LJ model $(q_c=0$),
dash-dotted curves are based on Eq.~(\ref{eq9}), and full curves on
Eq.~(\ref{eq11}).
The triangle shows the MC result for the critical point.
}
\end{figure}

\begin{figure}[h]\caption{\label{fig16}(Color online) Isothermal 
slice through the
phase diagram of the CO$_2$ + C$_5$H$_{12}$ system at $T=423.48
K$ (a) and $T=344.34 K$ (b). Full dots represent experimental data,\cite{80m}
 asterisks our Monte Carlo results for the model,
Eq.~(\ref{eq10}), while the full curve is the corresponding TPT1-MSA
prediction. The broken curve shows the corresponding TPT1-MSA result
for a CO$_2$ model with no quadrupole moment $(q_c=0)$.
Triangles are MC results for the critical points.
}
\end{figure}

\begin{figure}[h]\caption{\label{fig17}(Color online) Isothermal slice through the
phase diagram of the CO$_2$+C$_{16}$H$_{34}$ system at $T=486 K$,
showing MC results for the present model (open circles) and
comparing them to the results of the previous simulations
\cite{44} with $q_c=0, \xi=1$ (full dots) and $q_c=0, \xi = 0.886$
(asterisks). Squares show two sets of experimental data \cite{80} at two
temperatures that bracket the temperature used in the simulation.
Triangles are MC results for the critical point.
}
\end{figure}

\begin{figure}[h]\caption{\label{fig18}
(Color online) Critical line of the CO$_2$ +
C$_{16}$H$_{34}$ mixture, projected onto the $p^*,T^*$ plane
(pressure p and temperature T are rescaled with the LJ parameters
of the effective monomers of hexadecane as usual, $p^*=p \epsilon
/\sigma ^3$ and $T^*=k_BT/\epsilon$). Different symbols (as
indicated in the figure) denote data with $q_c=0, \xi=0.886$ (top
curve) and $\xi =0.9,\xi = 1$ (lowest curve), as well as data for
nonzero quadrupole moment, $q_c=0.387$ and $q=0.47$,
respectively.
}
\end{figure}

\begin{figure}[h]\caption{\label{fig10}(Color online) Final normalized order
parameter histogram $P(M)$ of Xenon-Krypton mixtures (curve) at
$T_1=228.78$. 
The simulated (s) and
reweighted (r) parameters (in units of $\epsilon _x$ and $\sigma_x$)
 are $\mu^*_{Ks}=-2.254$, $\mu^*_{Kr}=-2.2512$,
$\mu^*_{xs}=-3.972$, $\mu_{xr}^*=-3.9792$, $x_1=0.4$, $x_2 = 0.03$.
The dots show the universal 3d Ising model distribution.}
\end{figure}

\begin{figure}[h]\caption{\label{fig19} (Color online) (a) Plot 
of $B_4$ and $B_2$
as a function of $\mu_s $ for the CO$_2$ + C$_{16}$H$_{34}$
mixture (T=1.16). The chemical potential $\mu_p$ was always chosen such
that the equal weight rule was obeyed. Three different box linear
dimensions are included, as indicated. The broken horizontal lines
indicate the universal values $B_4^*$ and $B_2^*$ of the Ising
model at criticality, where the intersections of the curves for
$B_4$ and $B_2$ in the finite size scaling limit $(L \rightarrow
\infty)$ should occur. These data have been generated for the
mixing parameter $x_1=0.09$. The inset shows a plot of $B_4$ vs.
$x_1$ for $L=13.5 \sigma_p$ for three different values of $\mu_s$.
(b) Probability distribution $P(n_p,n_s)$ of the numbers of
polymers $(n_p)$ and solvent molecules $(n_s)$ for $L=13.5
\sigma_p$ at criticality.}
\end{figure}

\clearpage

\newpage
\clearpage

\begin{center}
\begin{table}
\vspace{0.5cm}
\begin{center}
\begin{tabular}{|c|c|c|c|c|c|c|c|}
\hline
 & q$_c$  & $\epsilon$/10$^{-21}$J  & $\sigma$/\AA & T$_c$/K & $\rho_c$/(mol/l) & p$_c$/bar & p$_{c,\mathrm{sim}}$/bar    \\
\hline
Kr & 0 & 2.8971(145) & 3.6524(126) & 209.46(2) & 11.0(1) & 55.20(6) & 52.33(66) \\
   & 0 & 2.8971(145) & 3.58782(2568) & 209.46(2) & 11.0(1) & 55.20(6) & 55.2(1.2) \\
\hline
Xe & 0 & 4.00747(2004) & 4.00053(685) & 289.74(2) & 8.37(1) & 58.42(6) & 55.08(48) 
\\
   & 0 & 4.00747(2004) & 3.92326(2803) & 289.74(2) & 8.37(1) & 58.42(6) & 58.4(1.3) \\
\hline
CO$_2$  &0&  4.20648(2104) & 3.69489(627)& 304.13(4)  &10.62(5)  &73.77(15) & 73.30(64) \\
 & 0.387 & 3.49047(1746) & 3.78467(641)& 304.13(4)  &10.62(5)  &73.77(15) &  73.10(64) 
\\
 & 0.47 &3.34887(1675) & 3.80341(645)& 304.13(4)  & 10.62(5) &73.77(15) & 73.10(64) 
\\
\hline
C$_6$H$_6$ & 0 & 7.77317(4041) & 5.16046(8863) & 562.0(8) &3.9(2) & 48.9(4) & 49.7(2.6) \\
     & 0.247 & 6.90972(3592) & 5.24125(9002) & 562.0(8) &3.9(2) & 48.9(4) & 49.5(2.6) \\
   & 0.382 & 6.47249(3365) & 5.28363(9075) & 562.0(8) &3.9(2) & 48.9(4) & 49.5(2.6)
\\
\hline
CH$_4$& 0 & 2.63624(1382) & 3.75782(2558)& 190.6(3)  & 10.1(2) & 46.1(3) & 43.70(95) \\
\hline
C$_3$H$_8$& 0 &5.11618(2573) & 4.71826(12360)& 369.9(2)  & 5.1(4) & 42.5(1) & 42.5(3.4)\\
\hline
C$_5$H$_{12}$& 0 &4.86594(2487)& 4.30303(3200)  & 469.8(5) & 3.22(7) & 33.6(6) & 31.90(75)\\
\hline
C$_{16}$H$_{34}$& 0 & 5.78879(4320)& 4.57052(4787)  & 722(4)  &0.967(30)  &14(2) & 12.80(42) \\
\hline
\end{tabular}
\end{center}
\end{table}
\end{center}
TABLE 1

\newpage
\clearpage
\begin{figure}
\includegraphics[angle=90,scale=0.8]{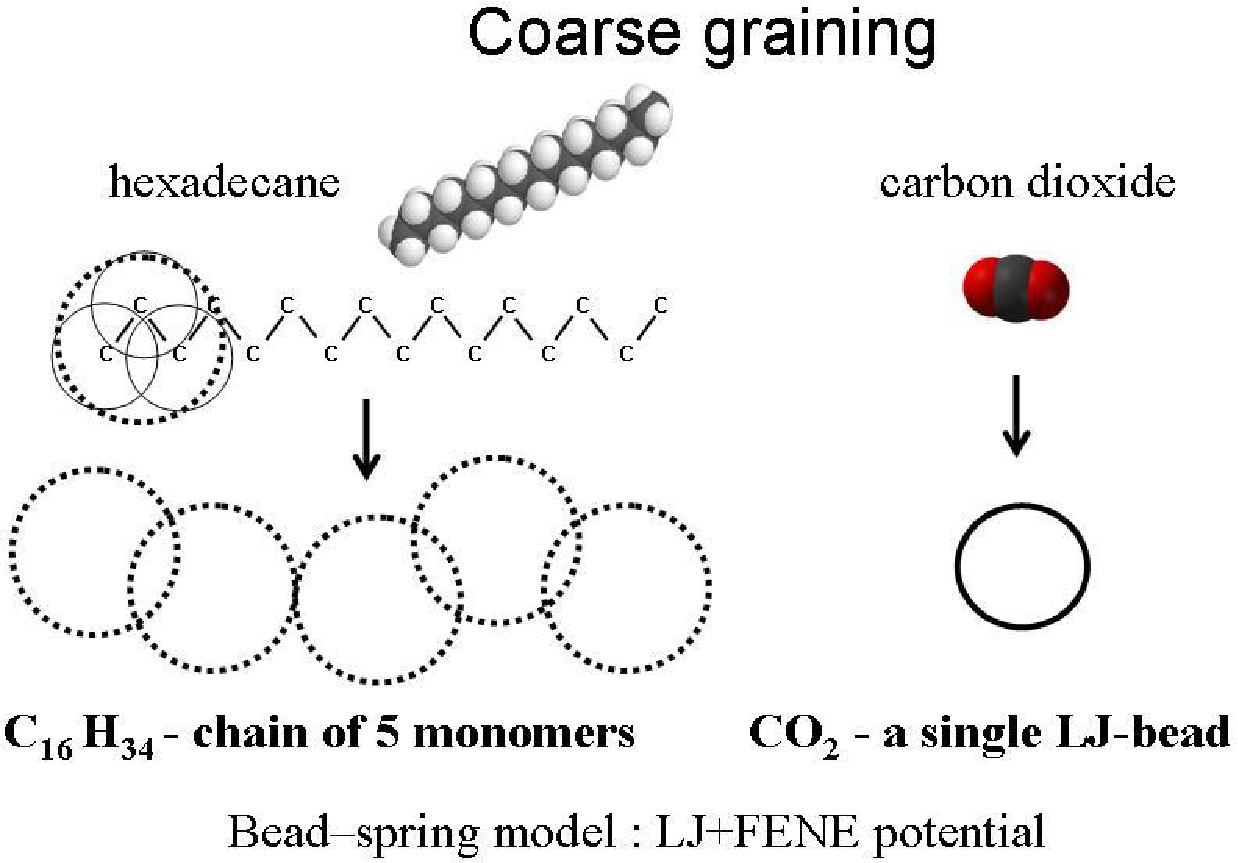}
\end{figure}
FIG.\ 1


\newpage
\clearpage
\begin{figure}
\includegraphics[angle=-90,scale=0.45]{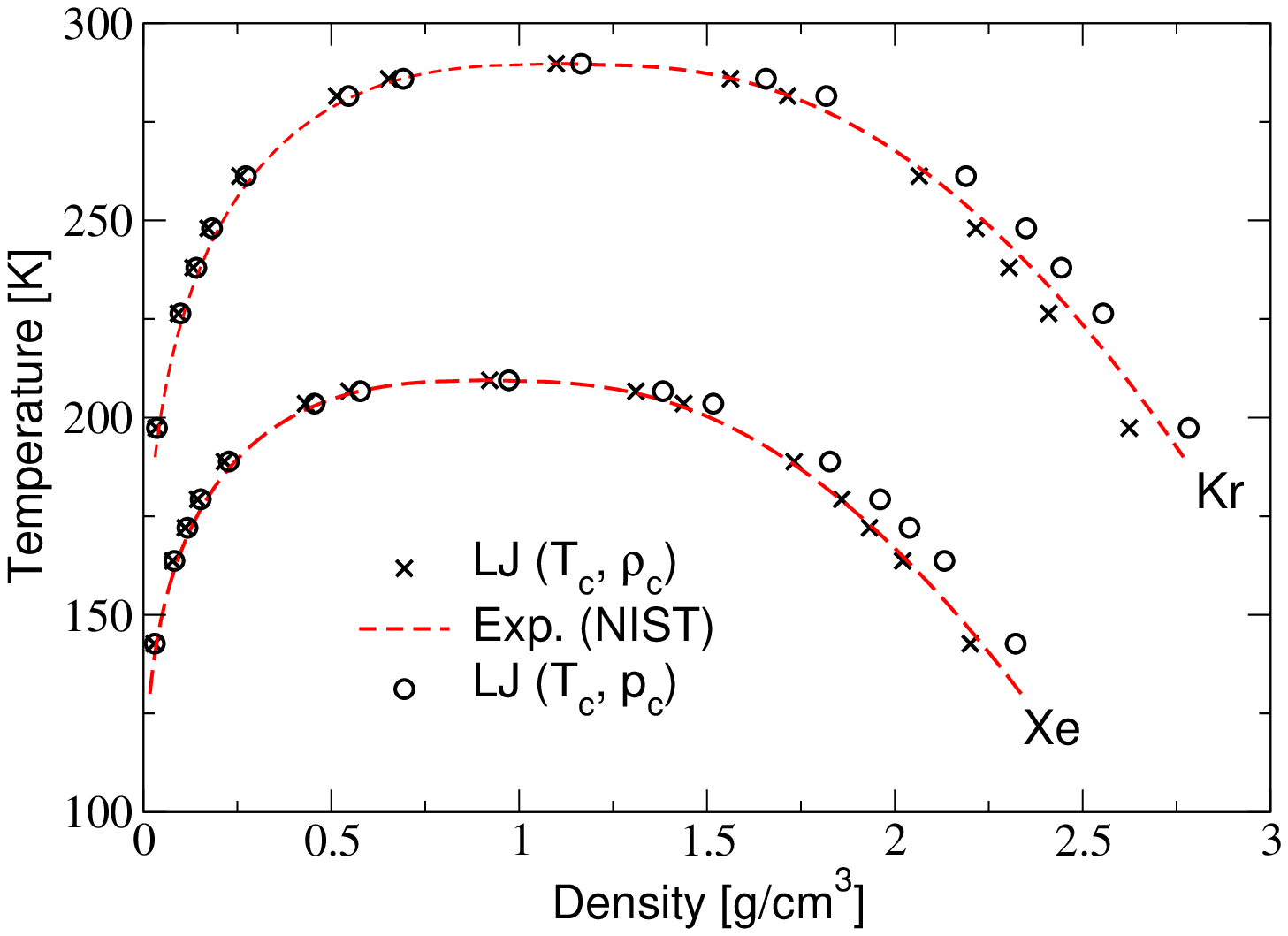}
\includegraphics[angle=-90,scale=0.45]{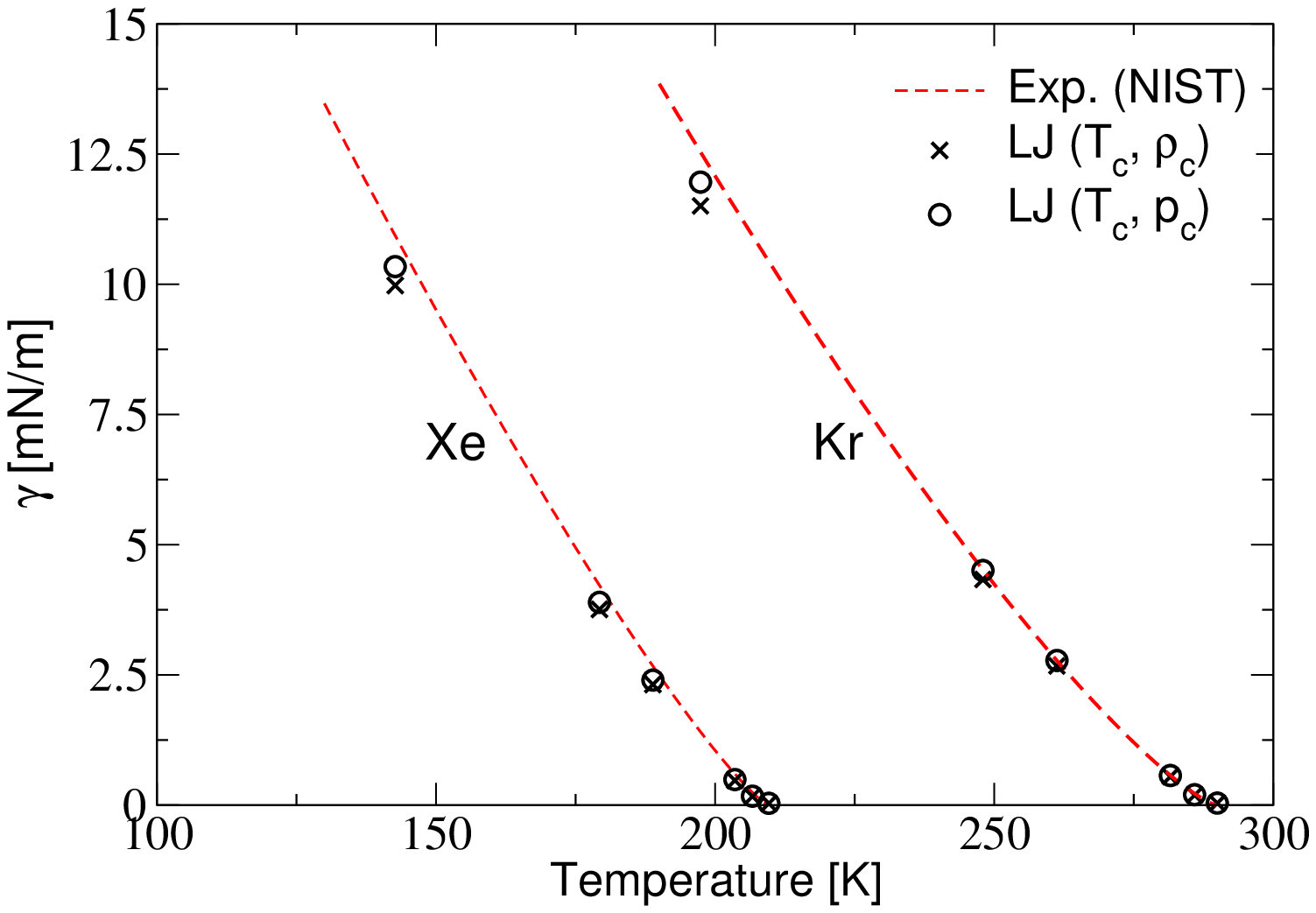}
\end{figure}
FIG.\ 2

\newpage
\clearpage
\begin{figure}
\includegraphics[angle=0,scale=0.7]{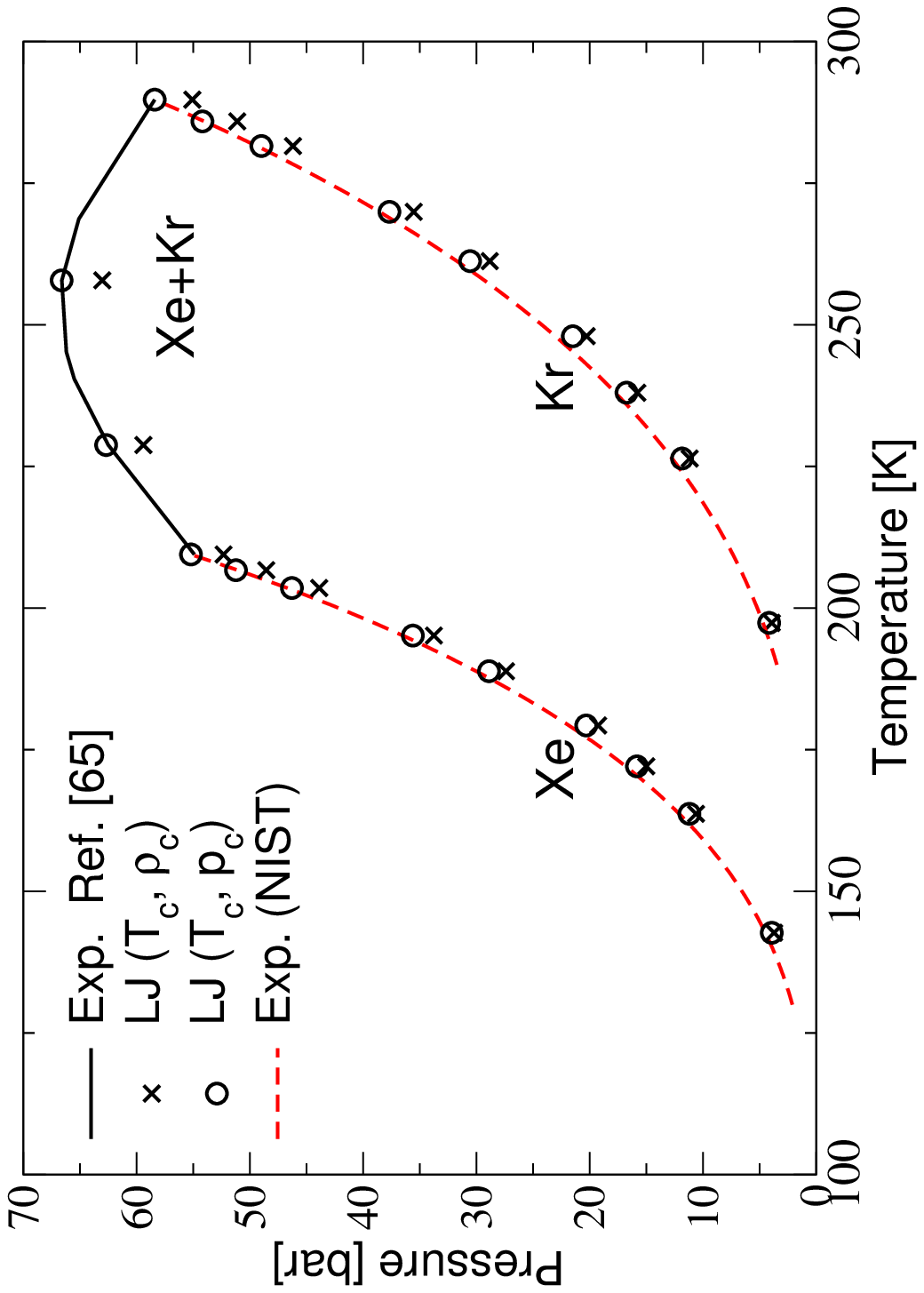}
\end{figure}
FIG.\ 3

\newpage
\clearpage
\begin{figure}
\includegraphics[angle=0,scale=0.67]{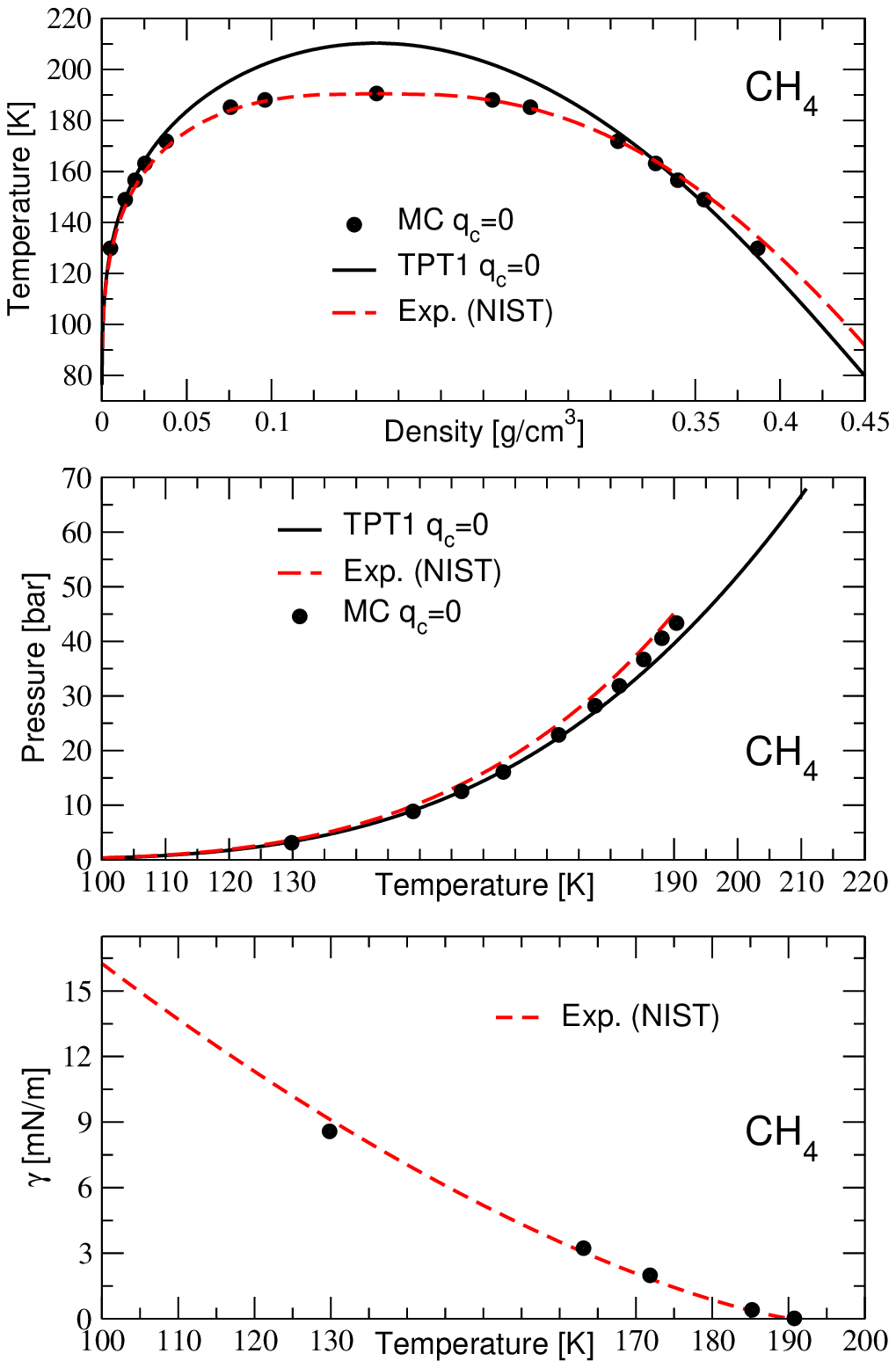}
\end{figure}
FIG.\ 4

\newpage
\clearpage
\begin{figure}
\includegraphics[angle=0,scale=0.67]{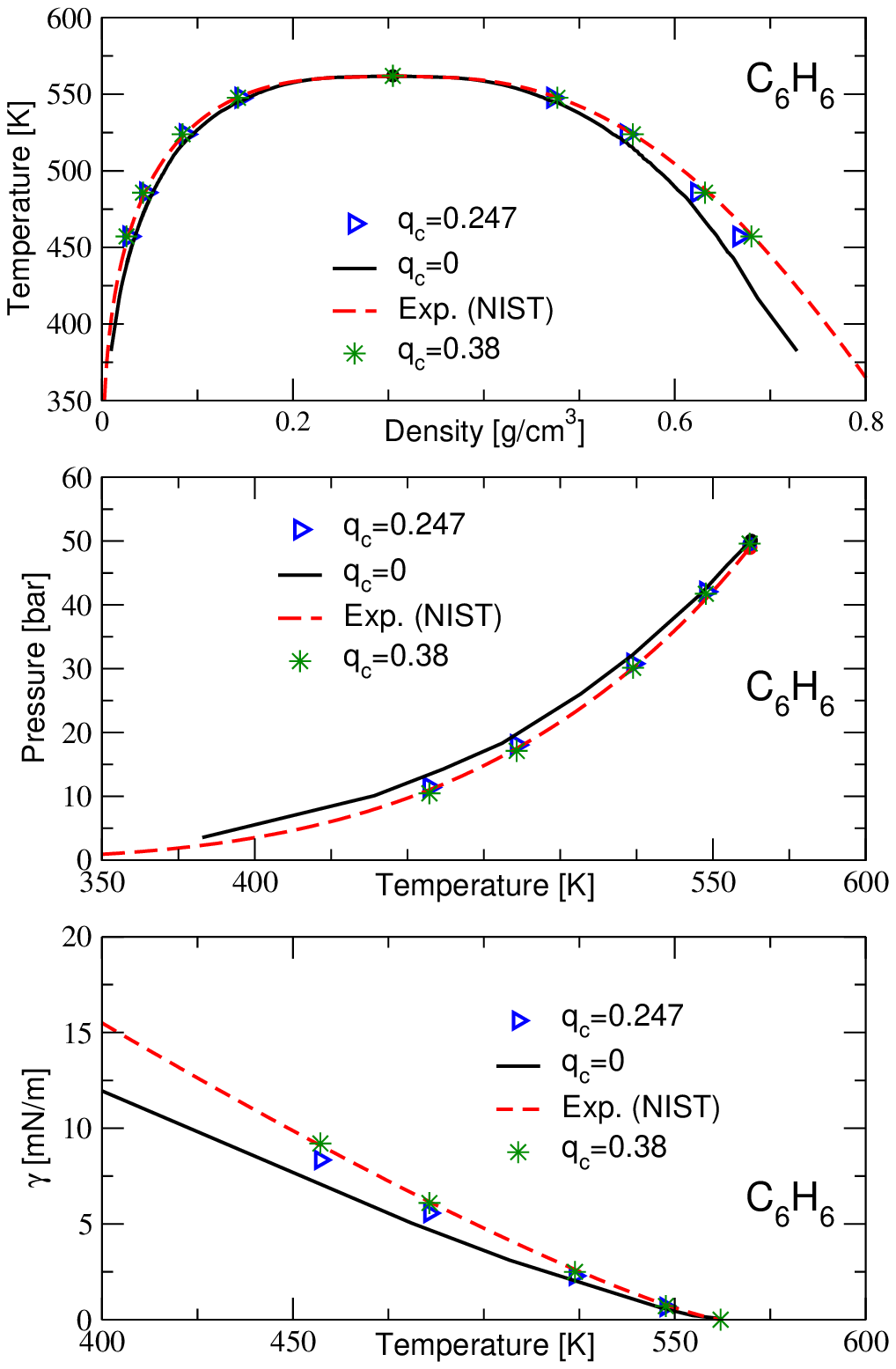}
\end{figure}
FIG.\ 5

\newpage
\clearpage
\begin{figure}
\includegraphics[angle=0,scale=0.67]{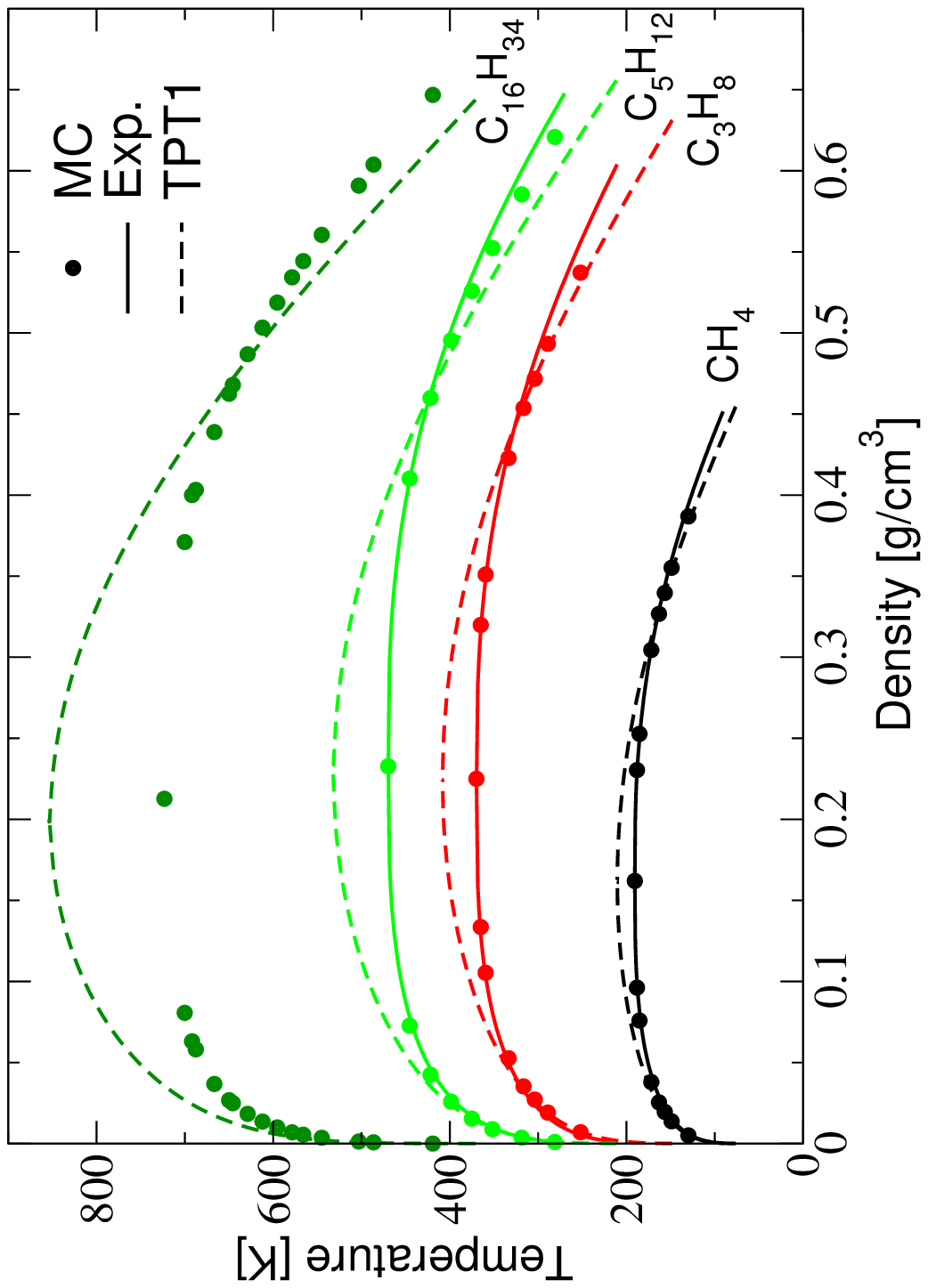}
\end{figure}
FIG.\ 6

\newpage
\clearpage
\begin{figure}
\includegraphics[angle=0,scale=0.67]{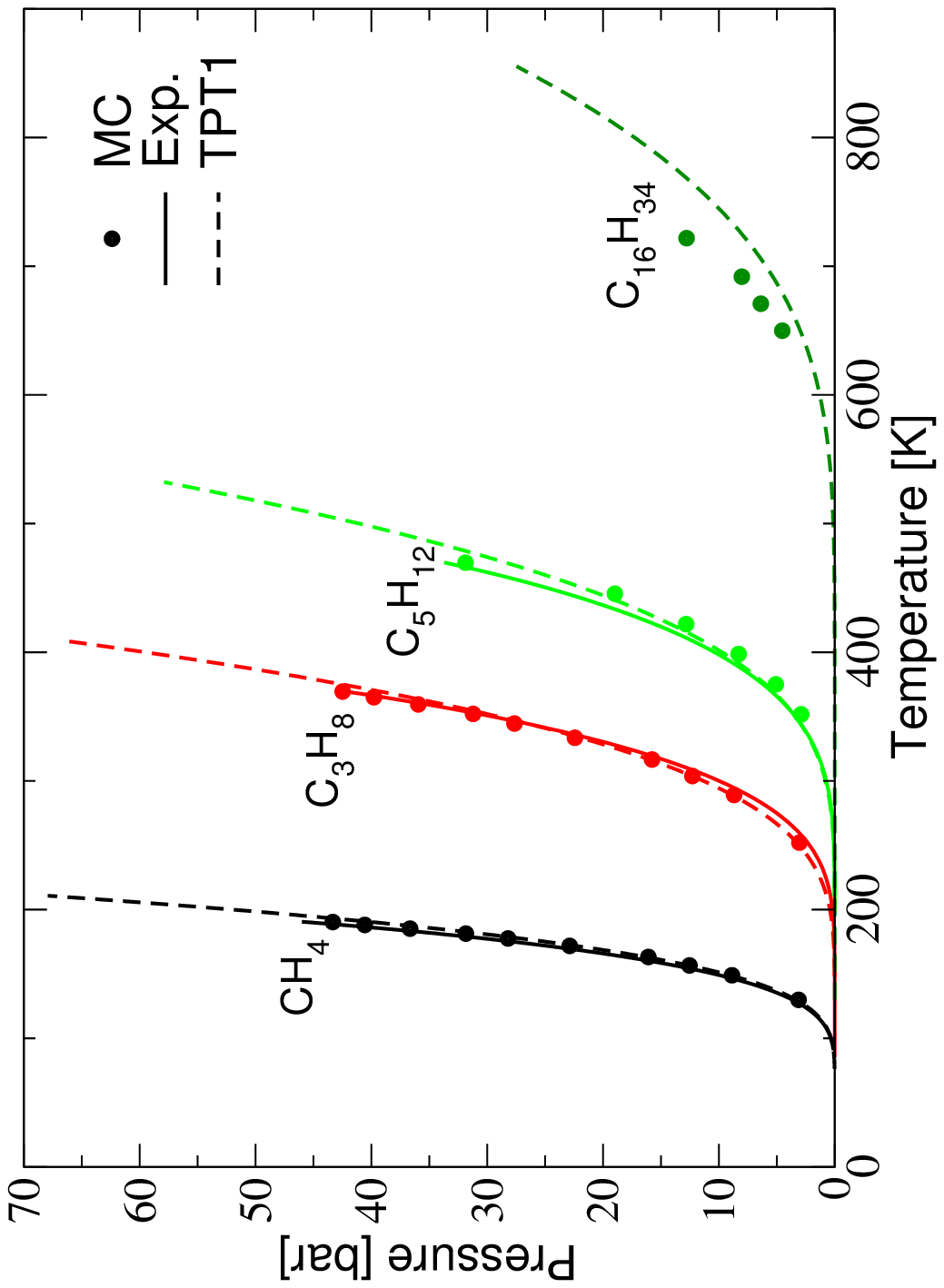}
\end{figure}
FIG.\ 7

\newpage
\clearpage
\begin{figure}
\includegraphics[angle=0,scale=0.67]{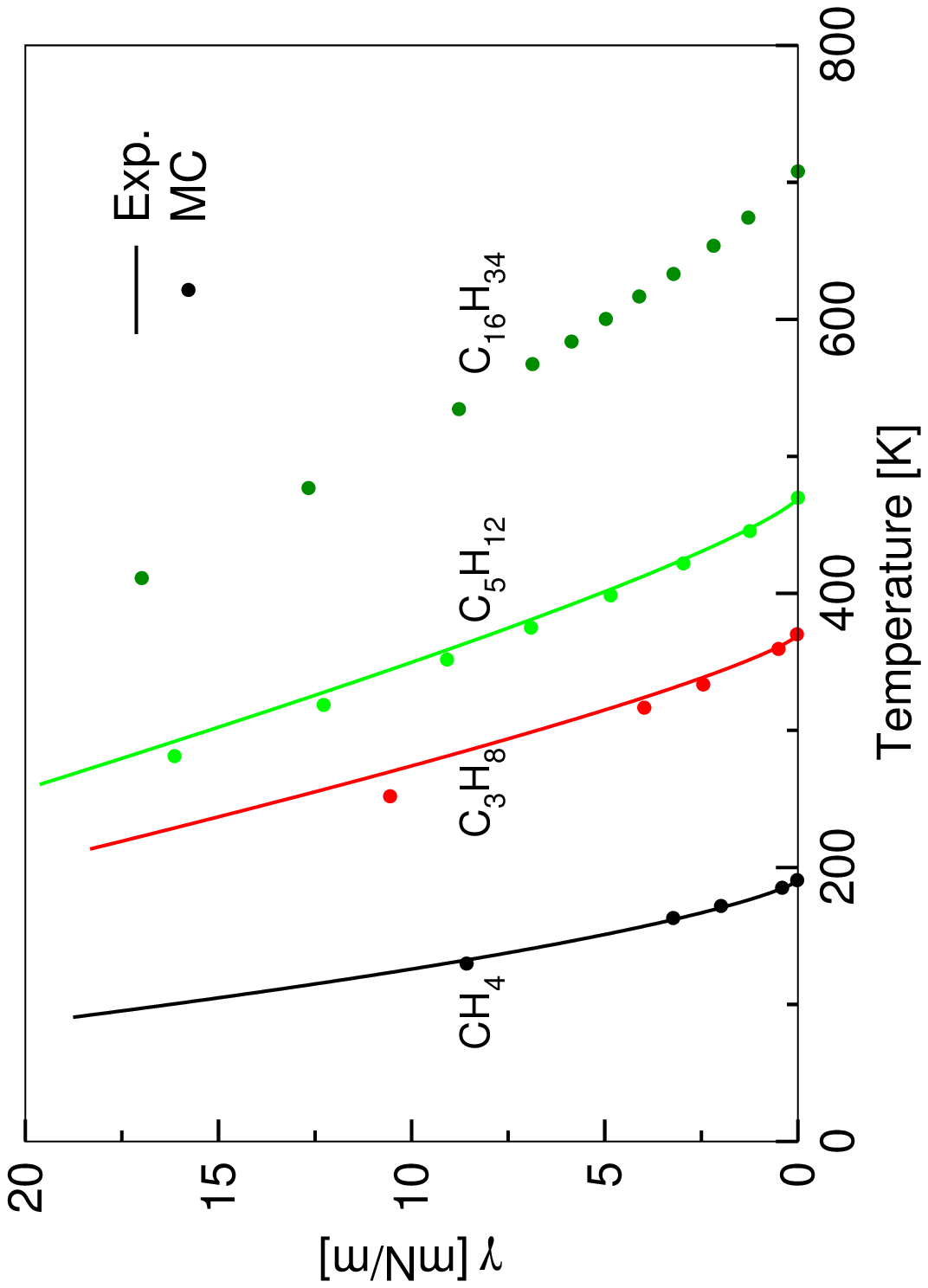}
\end{figure}
FIG.\ 8

\newpage
\clearpage
\begin{figure}
\includegraphics[angle=0,scale=0.7]{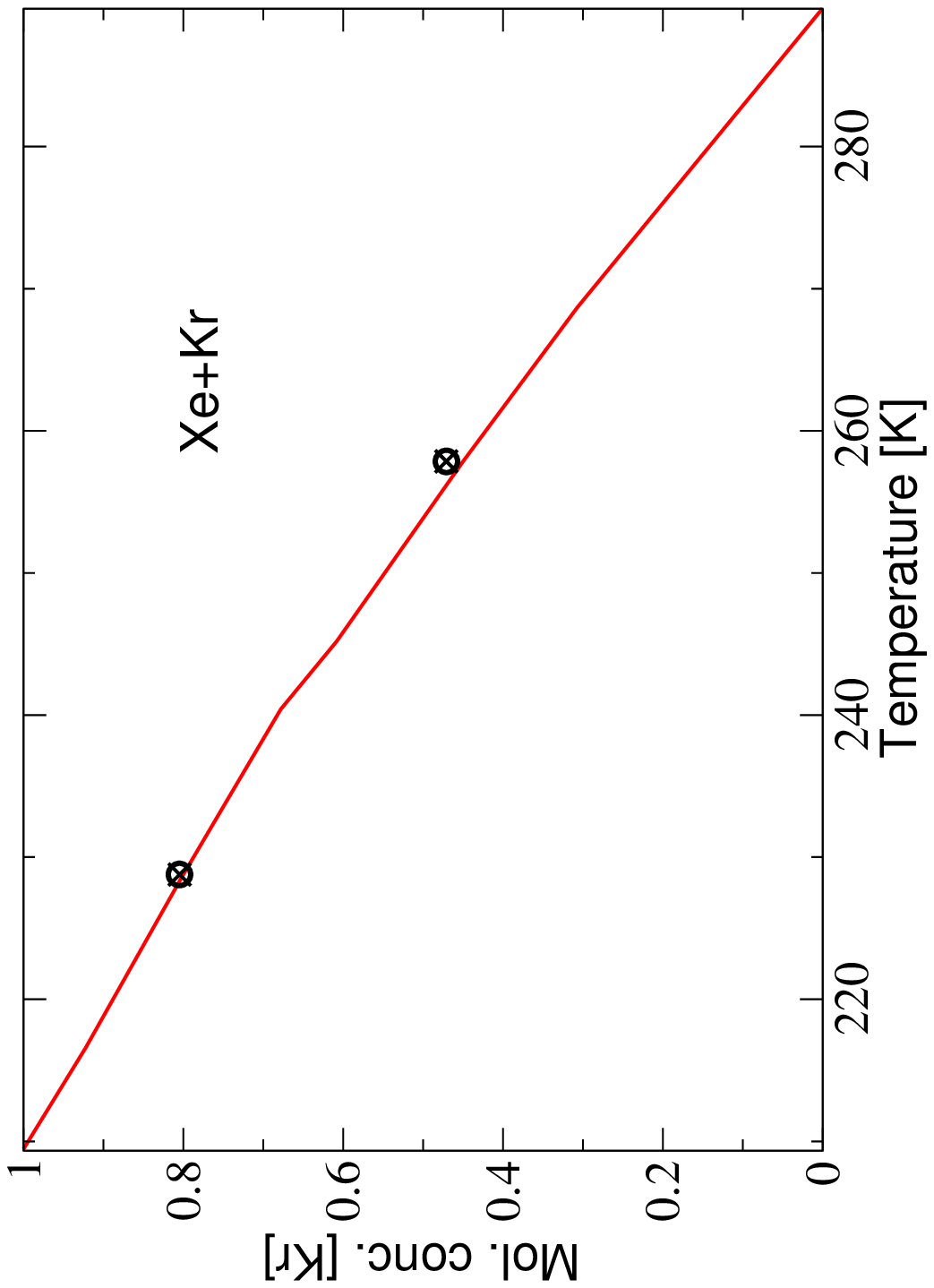}
\end{figure}
FIG.\ 9

\newpage
\clearpage
\begin{figure}
\includegraphics[angle=-90,scale=0.45]{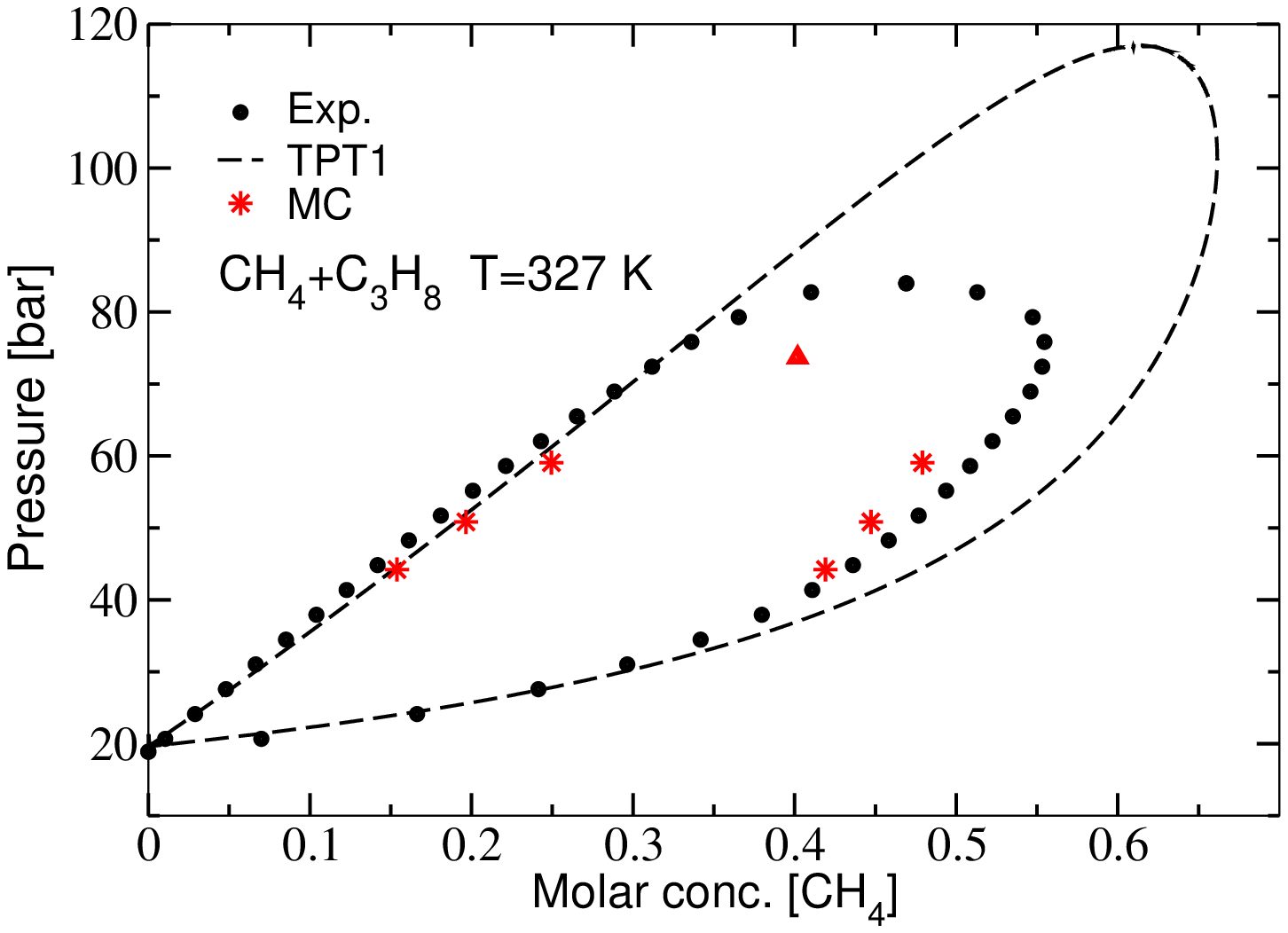}
\includegraphics[angle=-90,scale=0.45]{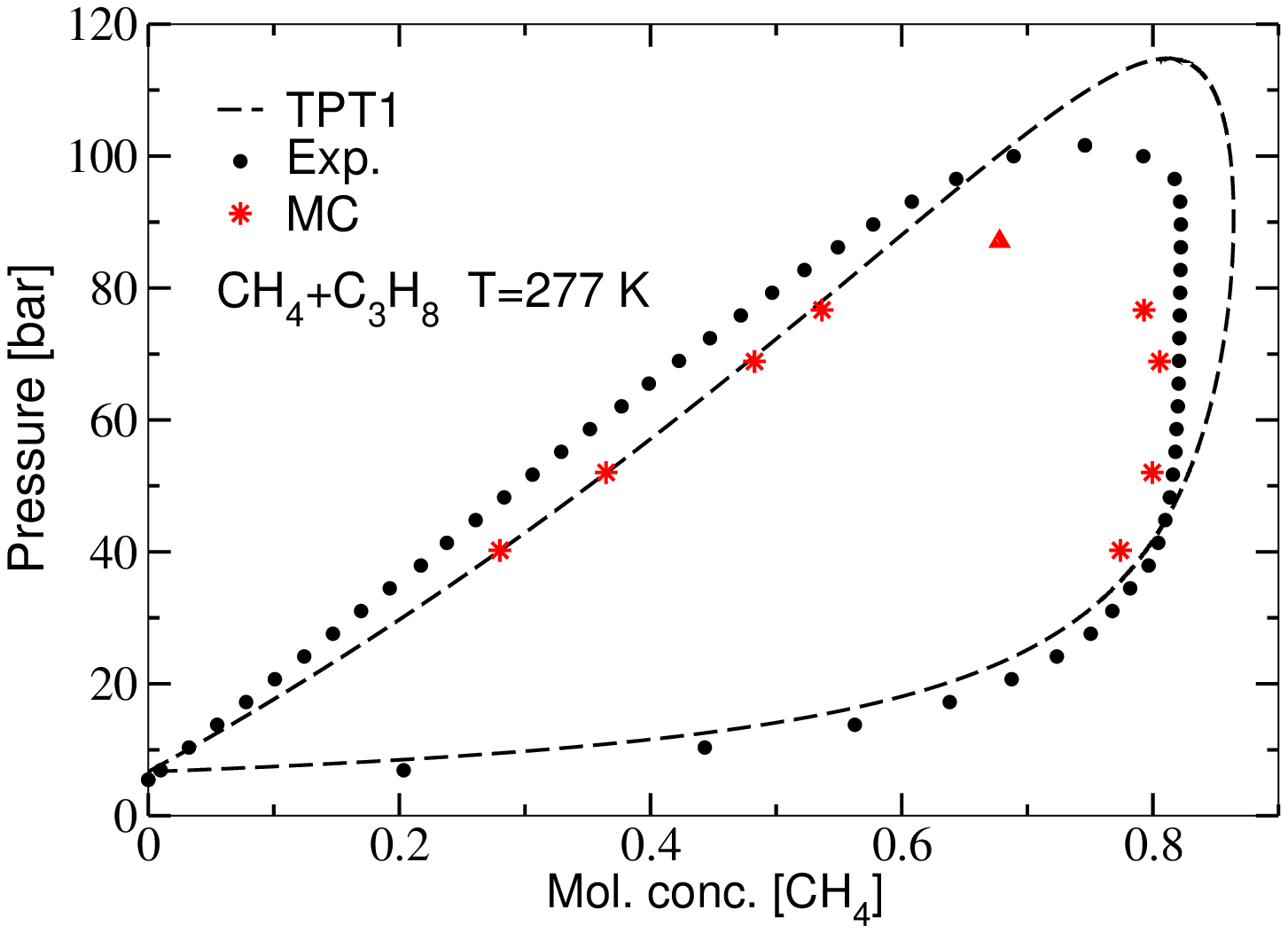}
\end{figure}
FIG.\ 10

\newpage
\clearpage
\begin{figure}
\includegraphics[angle=-90,scale=0.45]{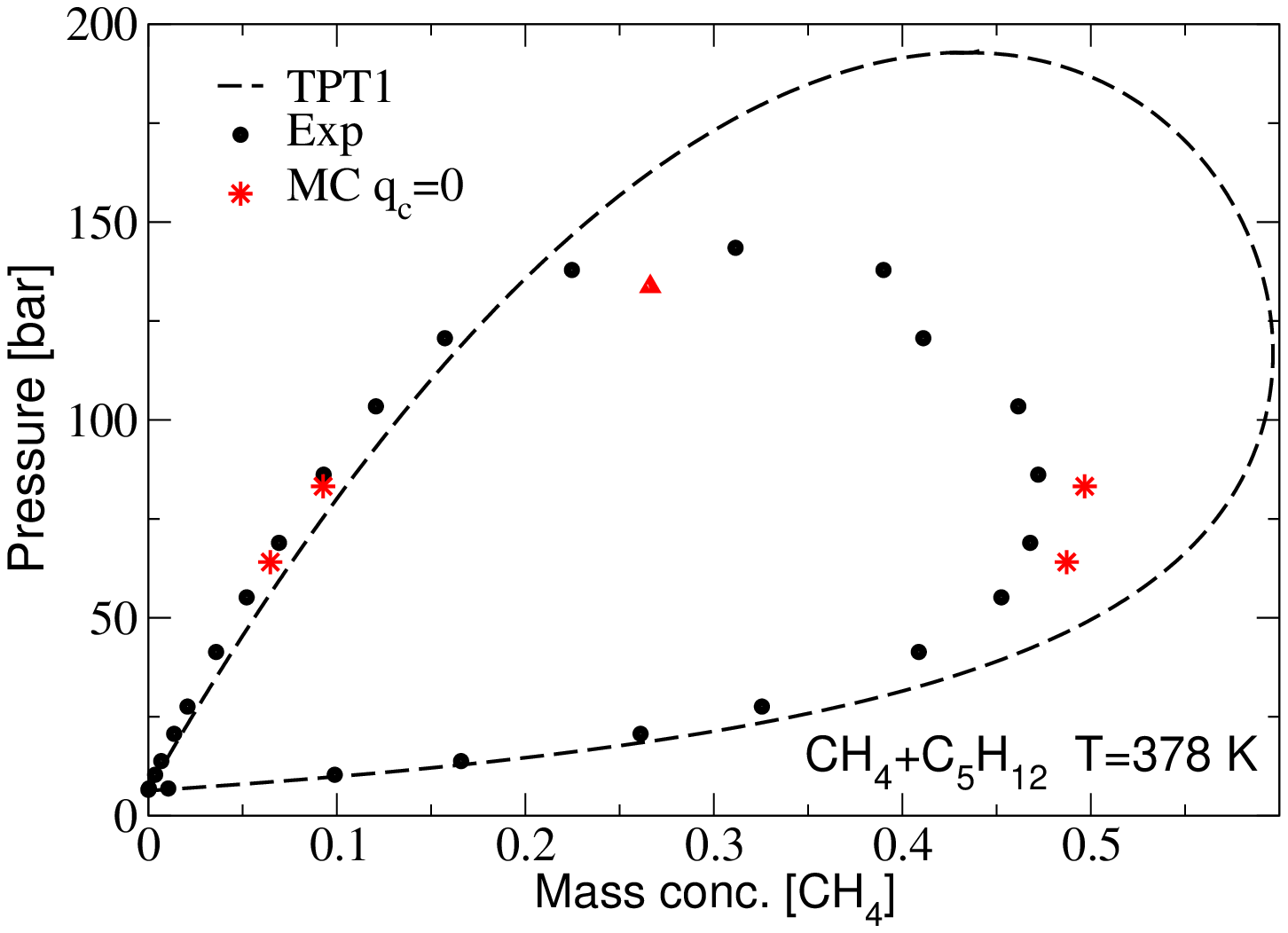}
\includegraphics[angle=-90,scale=0.45]{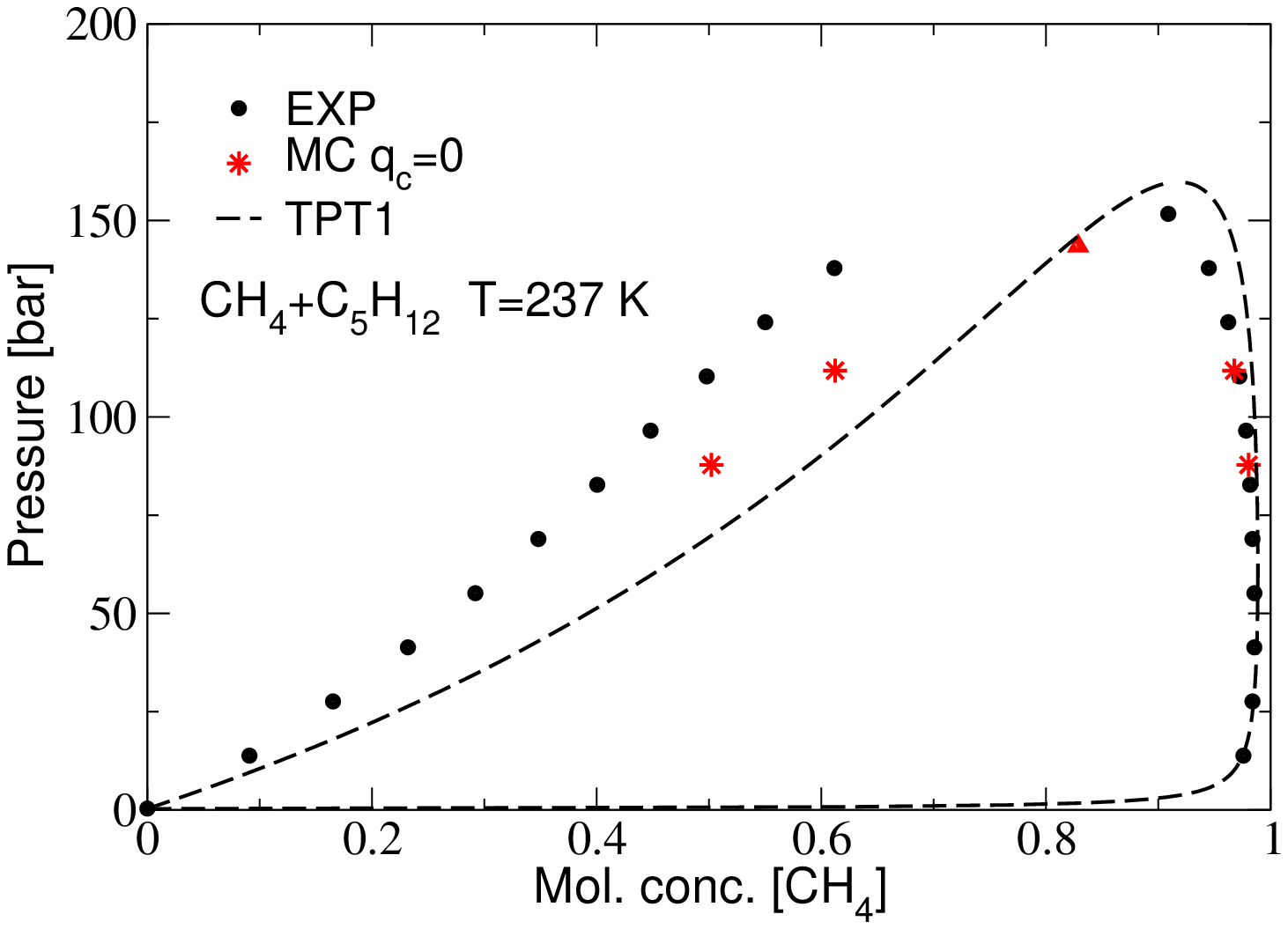}
\end{figure}
FIG.\ 11

\newpage
\clearpage
\begin{figure}
\includegraphics[angle=0,scale=0.67]{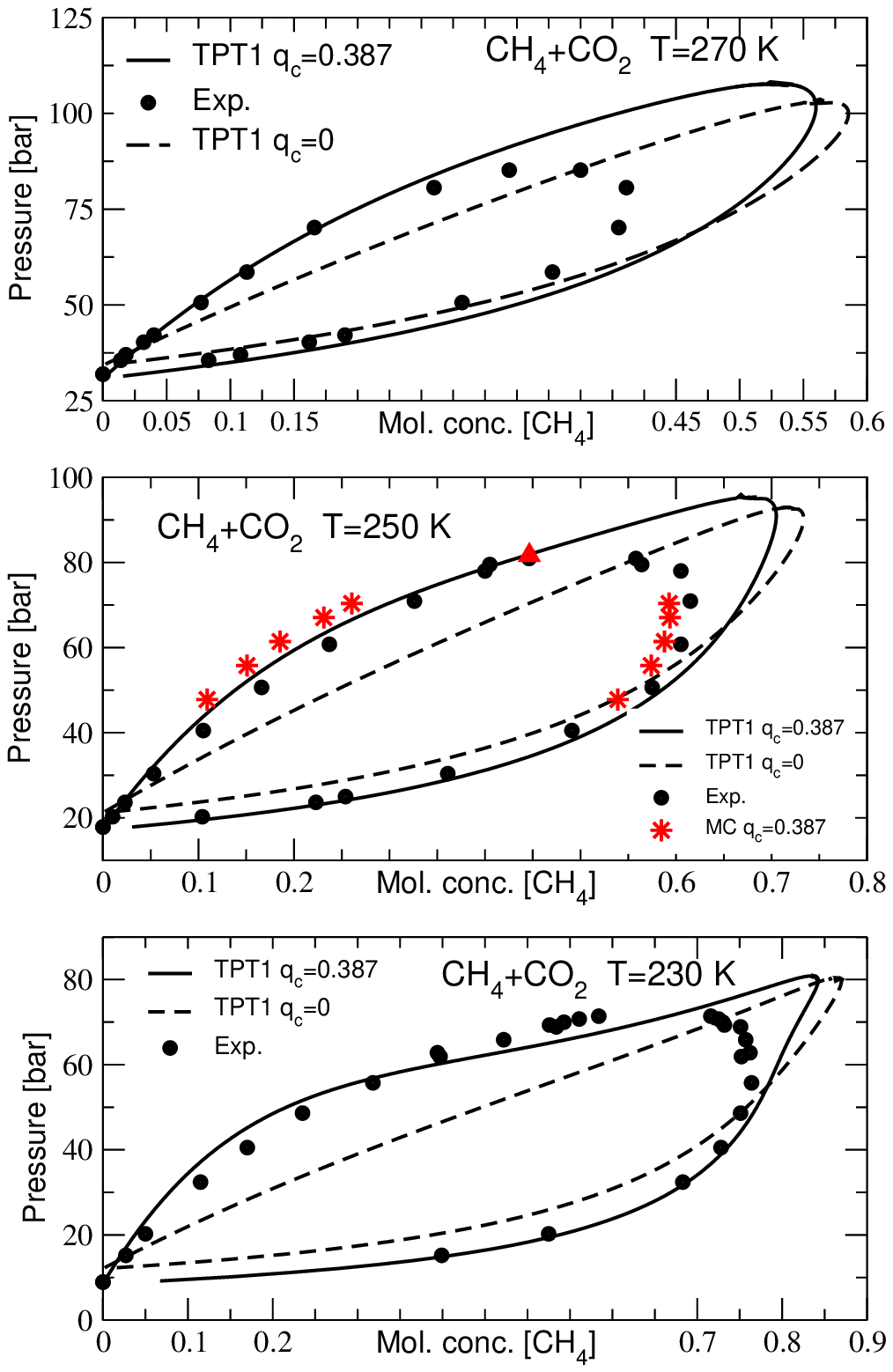}
\end{figure}
FIG.\ 12

\newpage
\clearpage
\begin{figure}
\includegraphics[angle=0,scale=0.67]{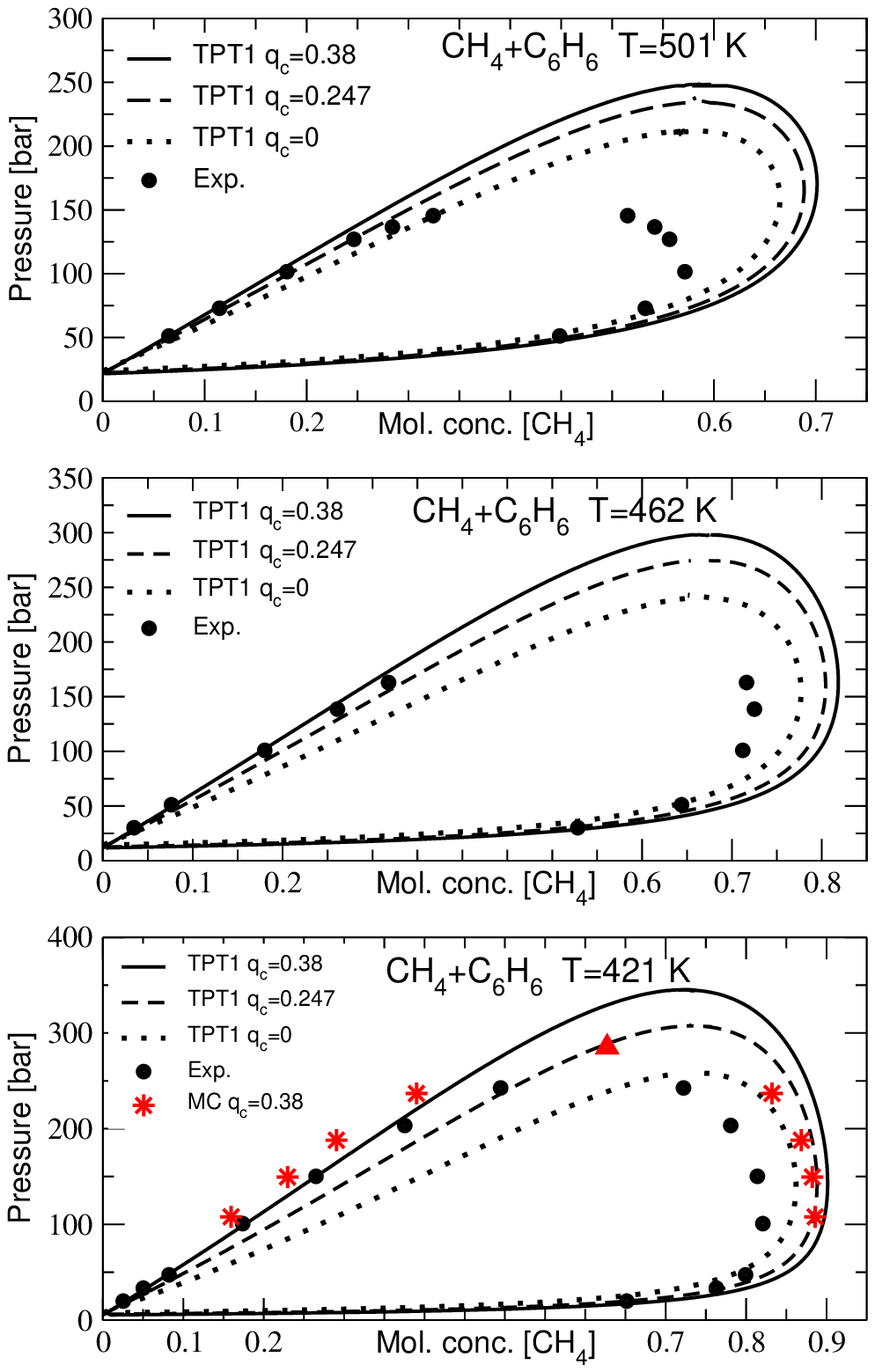}
\end{figure}
FIG.\ 13

\newpage
\clearpage
\begin{figure}
\includegraphics[angle=-90,scale=0.45]{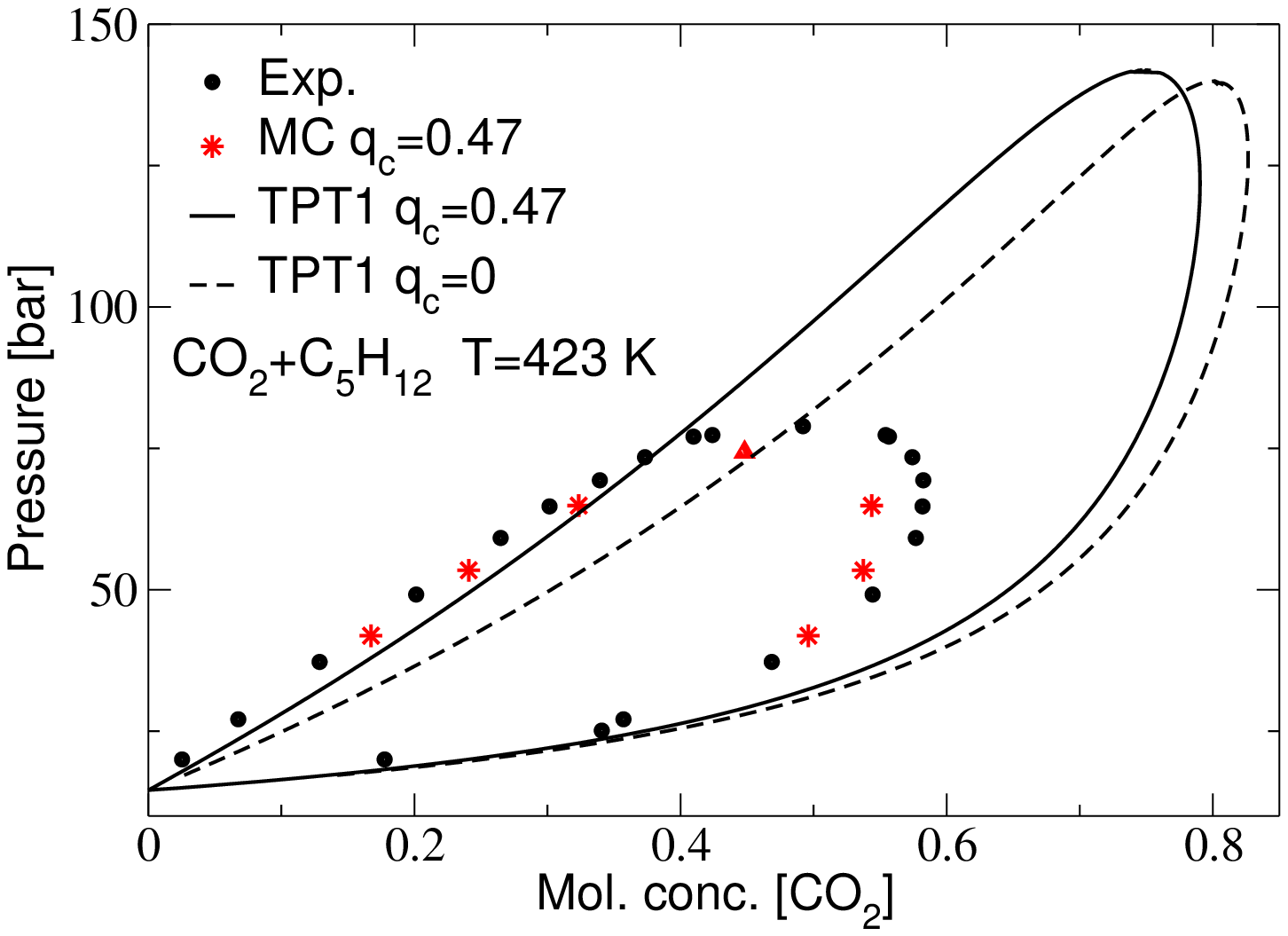}
\includegraphics[angle=-90,scale=0.45]{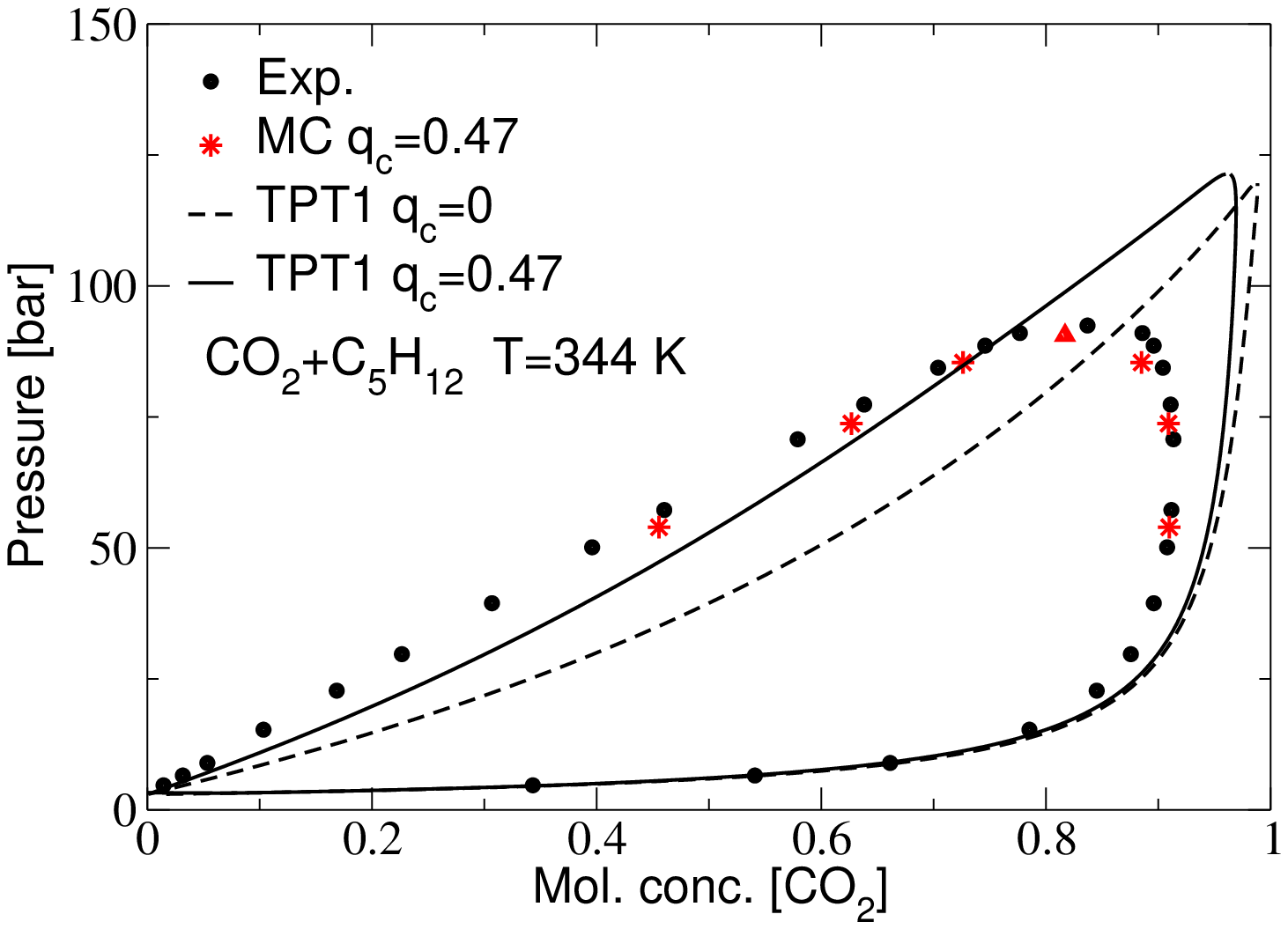}
\end{figure}
FIG.\ 14

\newpage
\clearpage
\begin{figure}
\includegraphics[angle=0,scale=0.7]{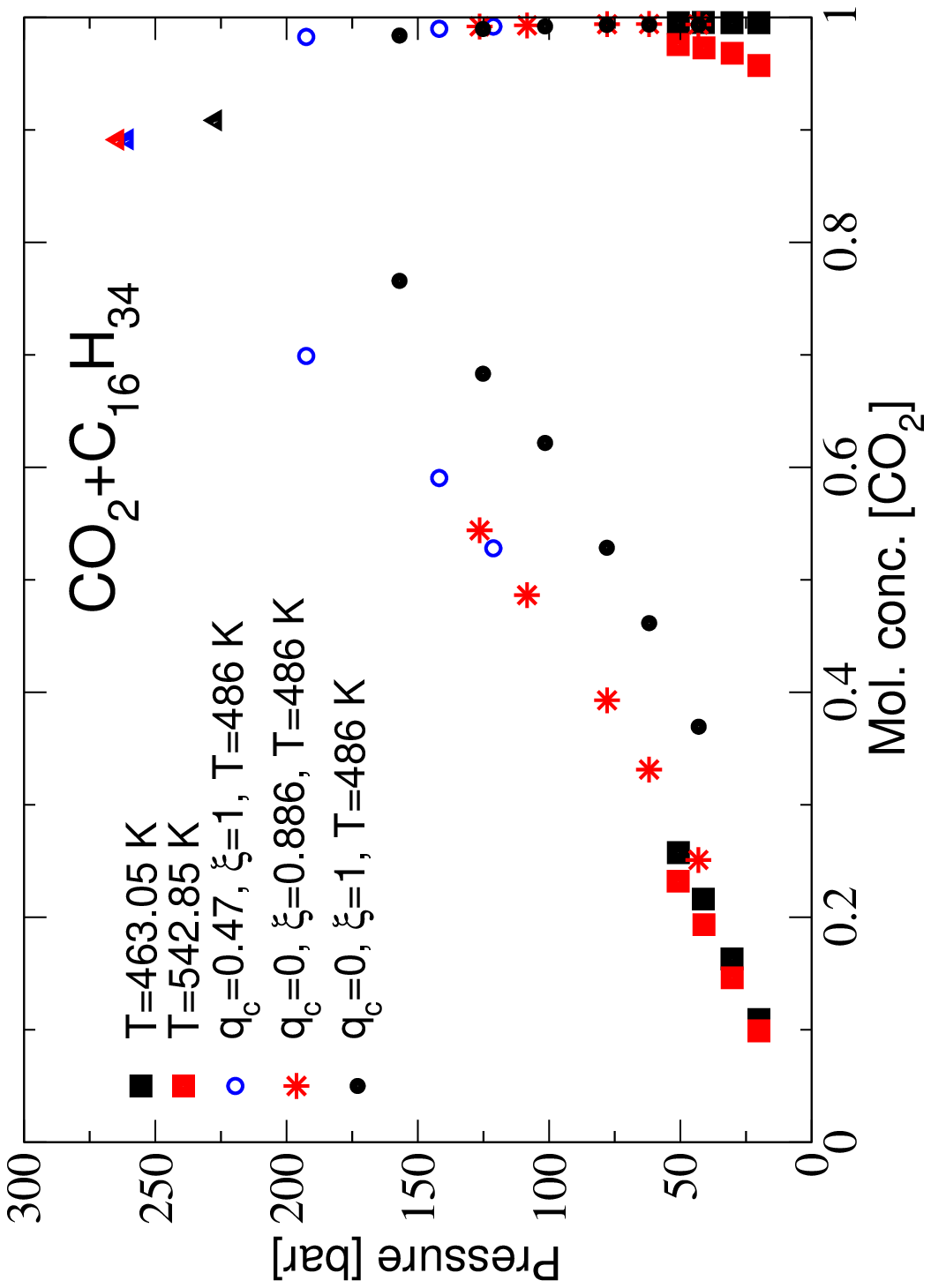}
\end{figure}
FIG.\ 15

\newpage
\clearpage
\begin{figure}
\includegraphics[angle=0,scale=0.7]{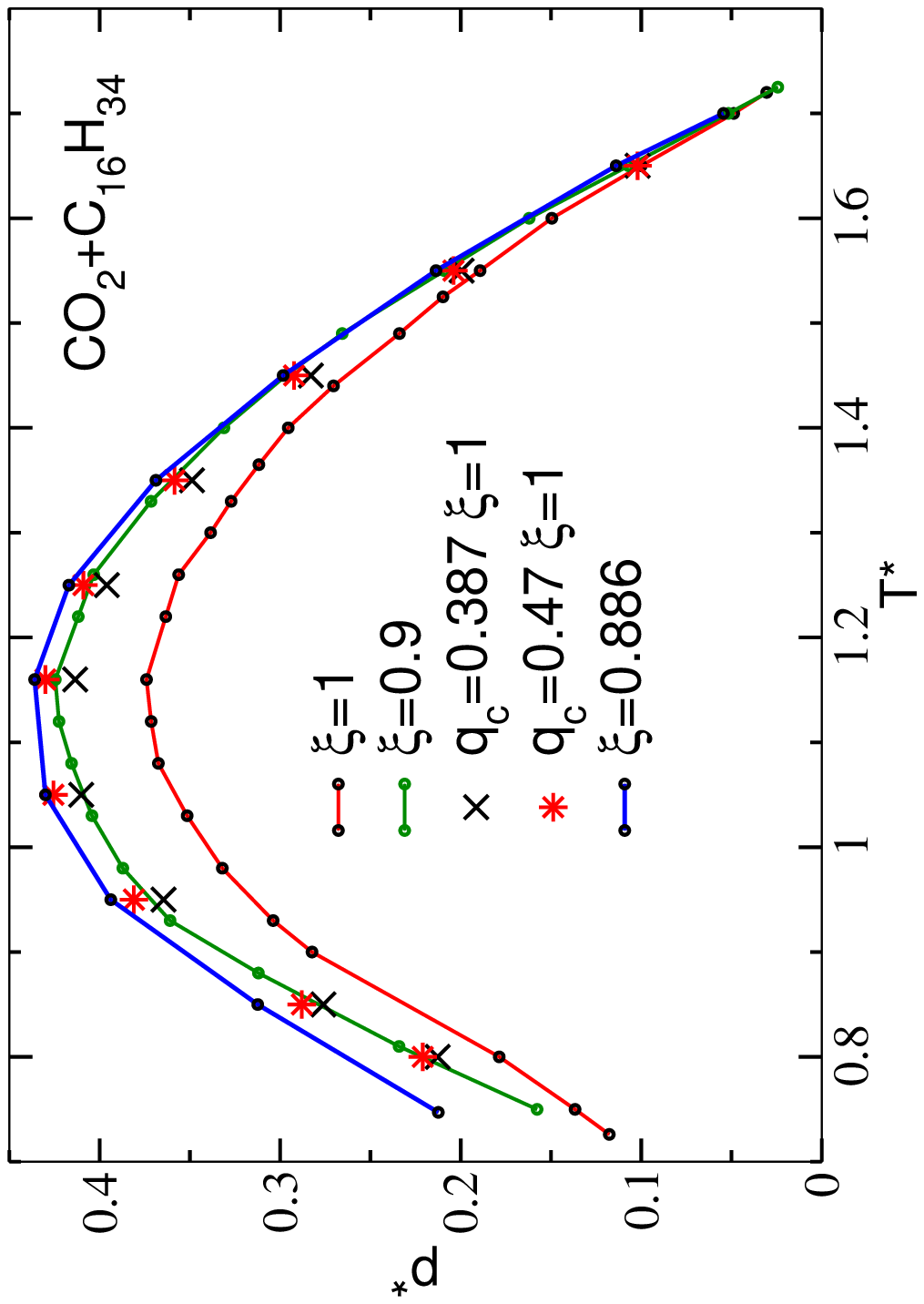}
\end{figure}
FIG.\ 16

\newpage
\clearpage
\begin{figure}
\includegraphics[angle=0,scale=0.67]{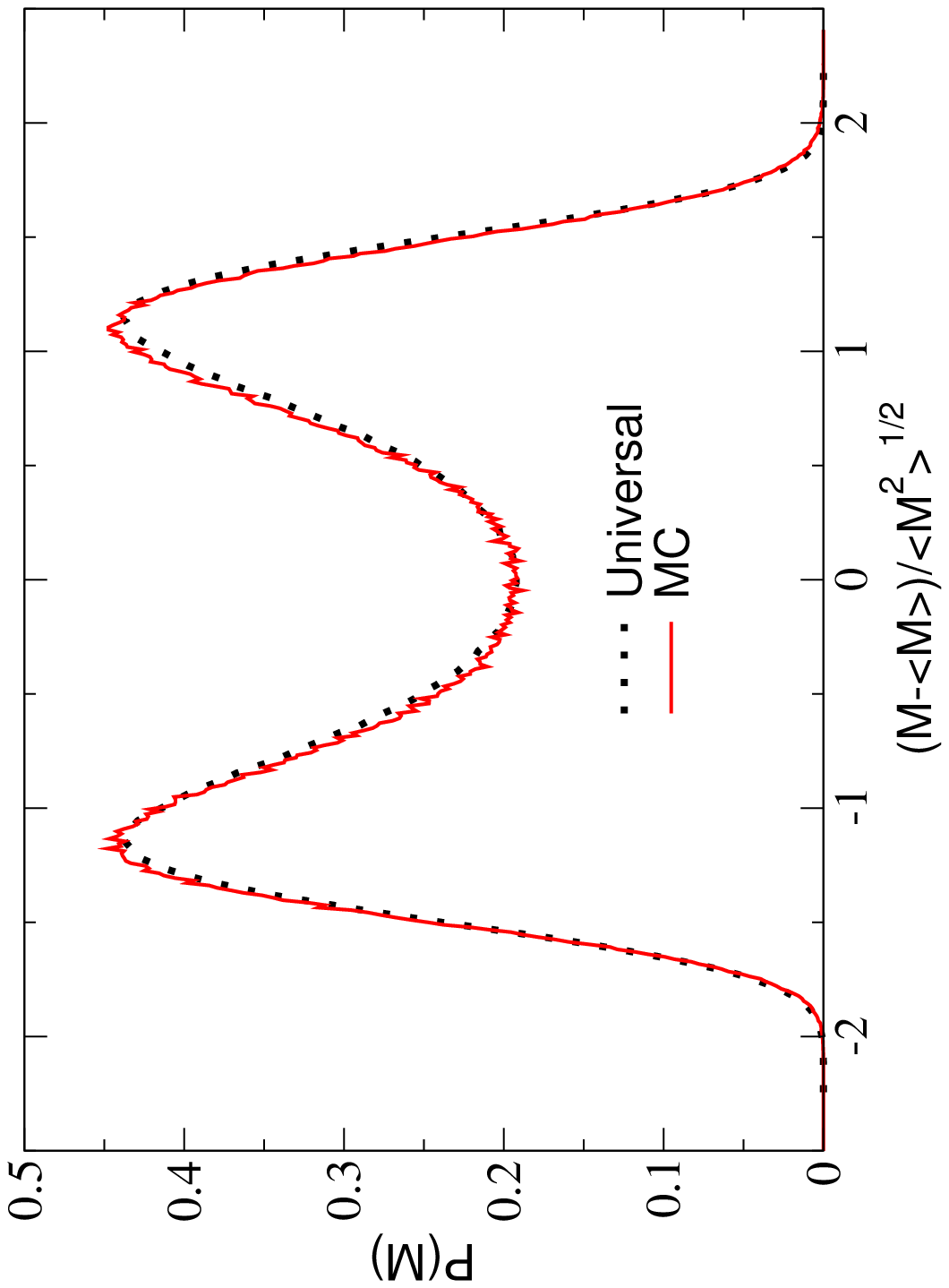}
\end{figure}
FIG.\ 17

\newpage
\clearpage
\begin{figure}
\includegraphics[angle=-90,scale=0.45]{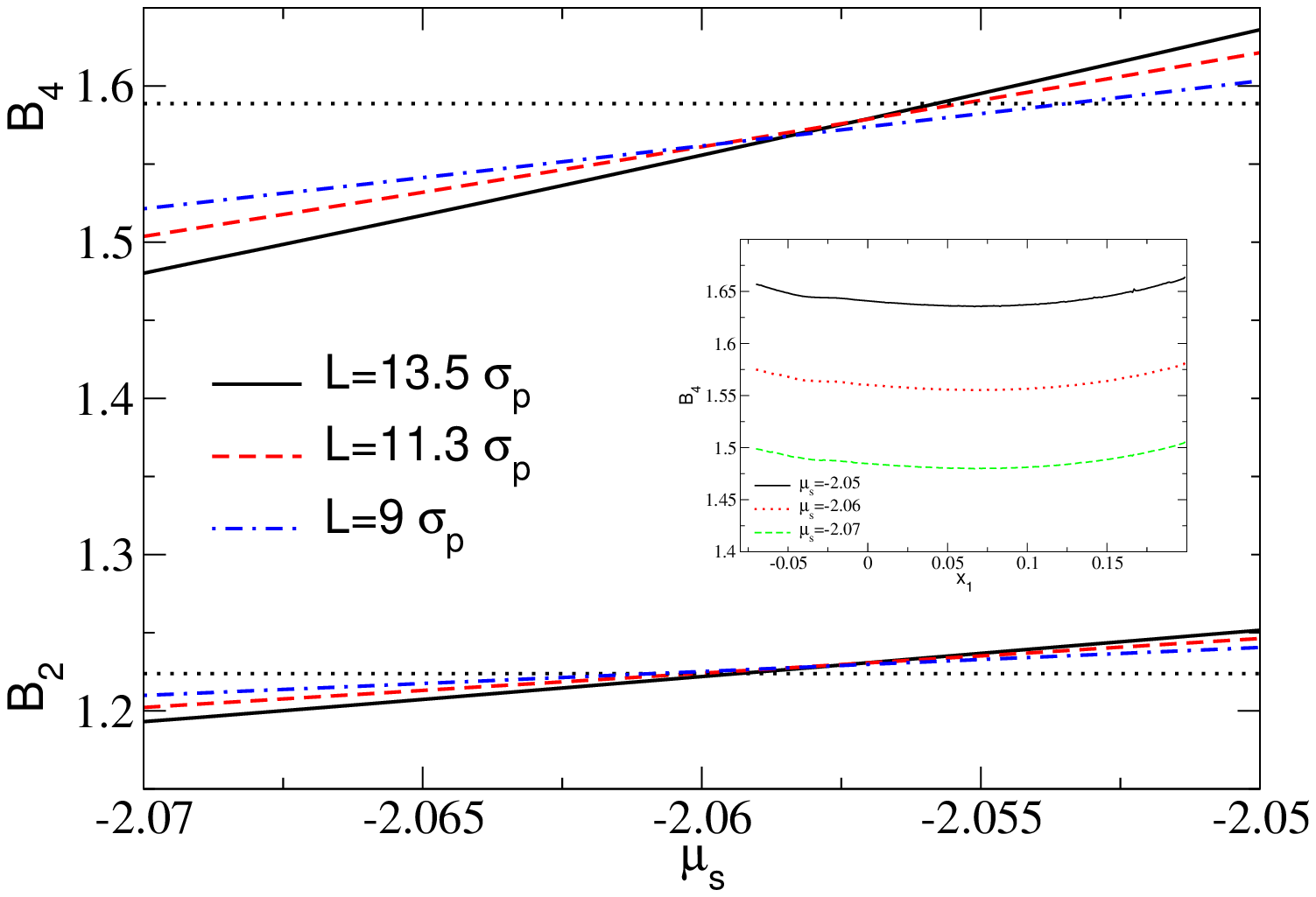}
\includegraphics[angle=0,scale=2]{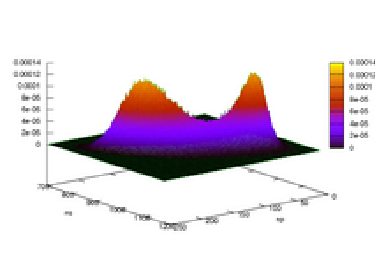}
\end{figure}
FIG.\ 18

\end{document}